\documentclass[graybox]{svmult}

\usepackage{mathptmx}       
\usepackage{helvet}         
\usepackage{courier}        
\usepackage[sort&compress,numbers]{natbib}
\usepackage{graphicx}        
\usepackage{multicol}        
\usepackage[bottom]{footmisc}
\usepackage[misc,geometry]{ifsym}

\usepackage{amsmath,amssymb,amsfonts,textcomp}

\usepackage{pdflscape}

\usepackage[caption=false]{subfig}
\usepackage[dvipsnames]{xcolor}

\usepackage{hyperref}

\begin{document}

\title*{Solving Nonlinear Parabolic Equations by a Strongly Implicit Finite-Difference Scheme}
\subtitle{Applications to the Finite-Speed Spreading of Non-Newtonian Viscous Gravity Currents}
\titlerunning{Numerical Simulation of Nonlinear Parabolic Equations}
\author{Aditya A.\ Ghodgaonkar and Ivan C.\ Christov}
\institute{Aditya A. Ghodgaonkar \at School of Mechanical Engineering, Purdue University, West Lafayette, Indiana 47907, USA
\and Ivan C. Christov (\Letter) \at School of Mechanical Engineering, Purdue University, West Lafayette, Indiana 47907, USA\\ e-mail  \href{mailto:christov@purdue.edu}{\texttt{christov@purdue.edu}}}

\maketitle


\abstract{We discuss the numerical solution of nonlinear parabolic partial differential equations, exhibiting finite speed of propagation, via a strongly implicit finite-difference scheme with formal truncation error $\mathcal{O}\left[(\Delta x)^2 + (\Delta t)^2 \right]$. Our application of interest is the spreading of viscous gravity currents in the study of which these type of differential equations arise. Viscous gravity currents are low Reynolds number (viscous forces dominate inertial forces) flow phenomena in which a dense, viscous fluid displaces a lighter (usually immiscible) fluid. The fluids may be confined by the sidewalls of a channel or propagate in an unconfined two-dimensional (or axisymmetric three-dimensional) geometry. Under the lubrication approximation, the mathematical description of the spreading of these fluids reduces to solving the so-called thin-film equation for the current's shape $h(x,t)$. To solve such nonlinear parabolic equations we propose a finite-difference scheme based on the Crank--Nicolson idea. We implement the scheme for problems involving a single spatial coordinate (i.e., two-dimensional, axisymmetric or spherically-symmetric three-dimensional currents) on an equispaced but staggered grid. We benchmark the scheme against analytical solutions and highlight its strong numerical stability by specifically considering the spreading of non-Newtonian power-law fluids in a variable-width confined channel-like geometry (a ``Hele-Shaw cell'') subject to a given mass conservation/balance constraint. We show that this constraint can be implemented by re-expressing it as  nonlinear flux boundary conditions on the domain's endpoints. Then, we show numerically that the scheme achieves its full second-order accuracy in space and time. We also highlight through numerical simulations how the proposed scheme accurately respects the mass conservation/balance constraint.}


\section{Introduction}
\label{sec:intro}

In his lucid 2015 book {\it Questions About Elastic Waves} \cite{Engl15}, Engelbrecht asks ``What is a wave?'' and answers ``As surprising as it may sound, there is no simple answer to this question.'' Indeed, the definition of `wave' depends on the physical context at hand \cite{Christov2014}. Although most wave phenomena in classical continuum mechanics are described by  hyperbolic (wave) equations, one of the surprises of 20\textsuperscript{th} century research into nonlinear partial differential equations (PDEs) is that \emph{certain} parabolic (diffusion) equations also yield structures with finite speed of propagation. Two examples are (i) a linear diffusion equation with a nonlinear reaction term \cite{KPP37,Fisher37}, and (ii) a diffusion equation that is nonlinear due to a concentration-dependent diffusivity \cite{Barenblatt1952}.\footnote{More specifically,  \citet{Barenblatt1952} (see also \cite[p.~13]{ZeldCollectedWorks}) credits the observation of finite-speed of propagation in a nonlinear diffusion equation to a difficult-to-find 1950 paper by Zeldovich and Kompaneets.} Indeed, it is known that certain aspects of wave phenomena can be reduced to a problem of solving a parabolic PDE, as gracefully illustrated by Engelbrecht \cite[Ch.~6]{Engl97} through a series of selected case studies; further examples include, but are not limited to: electromagnetic waves propagating along the earth's surface \cite{Vlasov1995}, seismic waves \cite{Jerzak2002}, underwater acoustics \cite{2011Jensen}, and the classical theory of nerve pulses \cite[\S6.4.2]{Engl97}, which nowadays has been updated by \citet{Engl18a} to a nonlinear hyperbolic (wave) model in the spirit of the Boussinesq paradigm \cite{Engl18b,CMP07}. 

Of special interest to the present discussion are physical problems that are modeled by nonlinear parabolic PDEs. These nonlinear problems lack general, all-encompassing solution methodologies. Instead, finding a solution often involves methods that are specific to the nature of the governing equation or the physical problem that it describes \cite[Ch.~4]{Evans2010} (see also the discussion in \cite{Christov2014} in the context of heat conduction). The classical examples of nonlinear parabolic PDEs admitting traveling wave solutions come from heat conduction \cite[Ch.~X]{ZR02} (see also \cite{Straughan2011}) and thermoelasticity \cite{Straughan2011,BerezVan2017}. The sense in which these nonlinear parabolic PDEs admit traveling-wave and `wavefront' solutions now rests upon solid mathematical foundations \cite{GildKers2004,Vazquez2007}, including the case of gradient-dependent nonlinearity \cite{Tedeev2015} (e.g., the last case in table~\ref{tb:eq_models} to be discussed below).

A classical example of a nonlinear parabolic PDE governing the finite-speed wave-like motion of a substance arises in the study of an ideal gas spreading in a uniform porous medium \cite{Barenblatt1952}. A similar nonlinear parabolic equation was derived for the interface between a viscous fluid spreading horizontally underneath another fluid of lower density ($\Delta \rho > 0$ between the fluids) \cite{Huppert1982a}. The motion of the denser fluid is dictated by a balance of buoyancy and viscous forces at a low Reynolds number (viscous forces dominate inertial forces). Such viscous gravity current flows are characterized by `slender' fluid profiles i.e., they have small aspect ratios ($h/L \ll 1$, where $h$ and $L$ are typical vertical and horizontal length scales, respectively). Therefore, these flows can be modeled by lubrication theory \cite[Ch.~6]{L07}. Generically, one obtains a nonlinear parabolic equation for the gravity current's shape $h$ as a function of the flow-wise coordinate $x$ and time $t$. The case of the spreading of a fixed mass of Newtonian fluid was originally explored contemporaneously by \citet{Didden1982} and \citet{Huppert1982a}.  

\begin{figure}
\subfloat[][]{\includegraphics[width=0.5\textwidth]{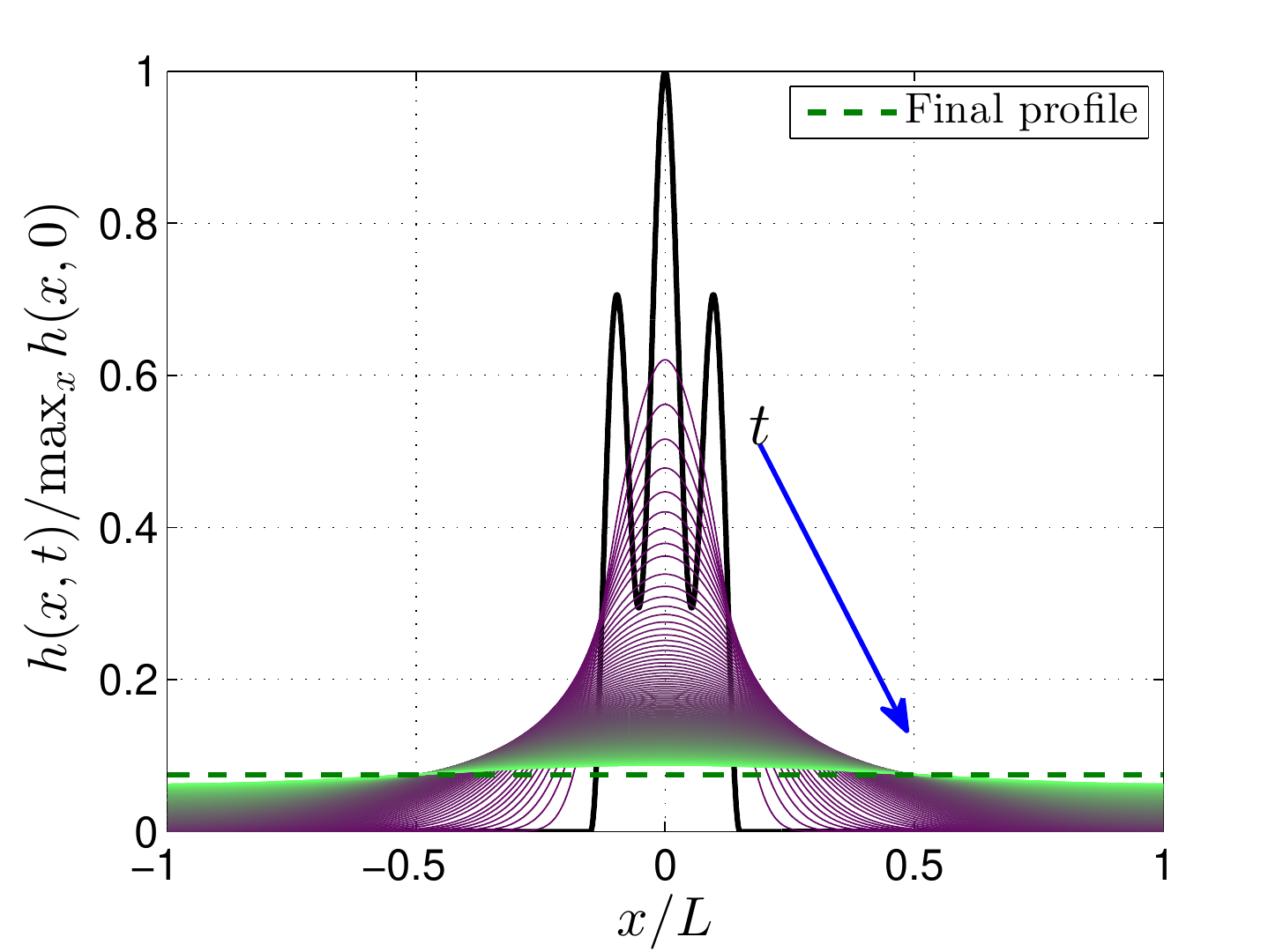}\label{fig:introH}}
\hfill
\subfloat[][]{\includegraphics[width=0.5\textwidth]{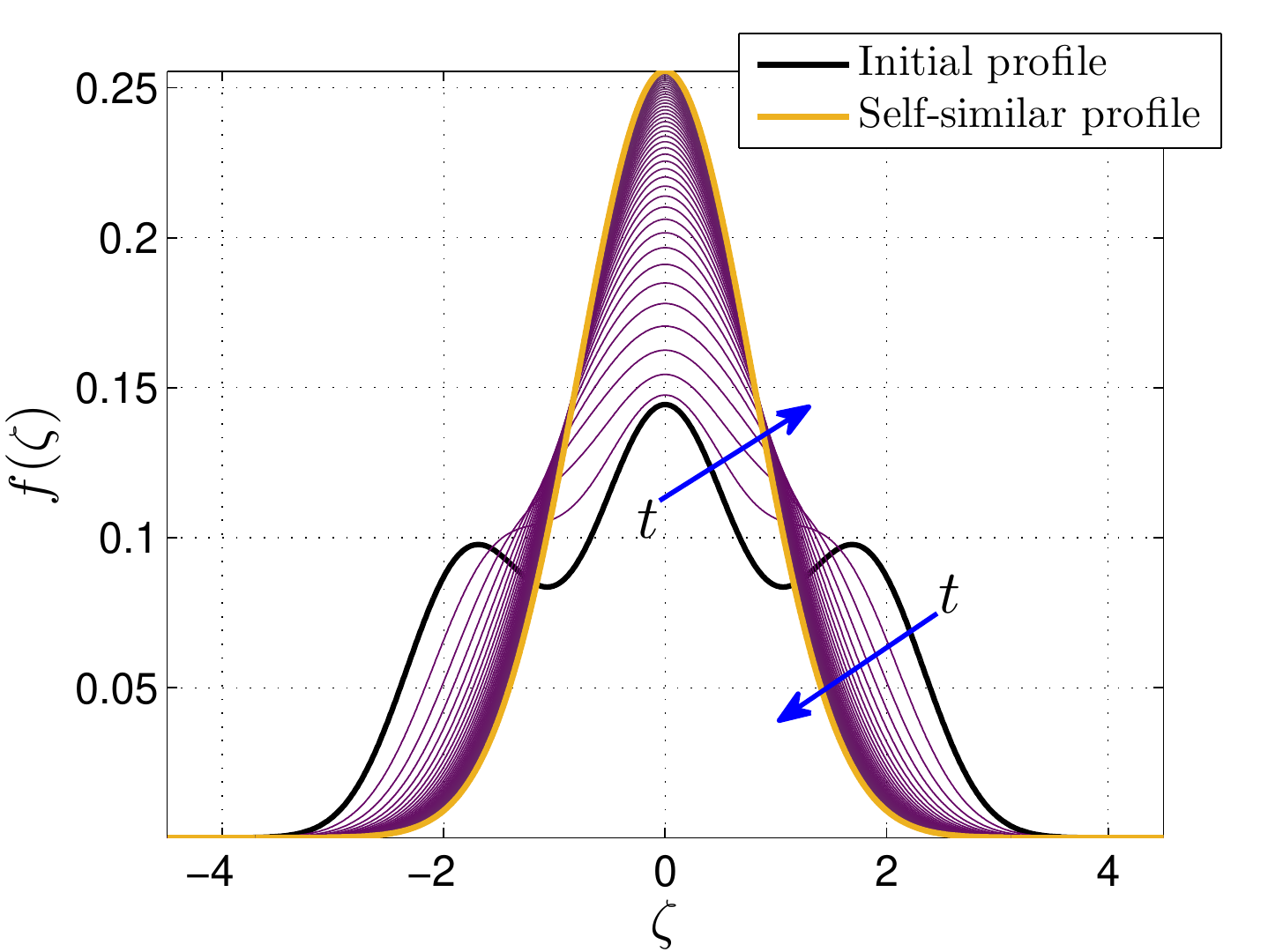}\label{fig:introF}}
\caption{Spreading via 1D linear diffusion and the approach to the universal intermediate self-similar asymptotics. (a) An arbitrary wavy IC $h(x,0)$ spreads and levels until reaching a flat steady state $h_\infty$. (b) A first-kind self-similar transformation (obtained from dimensional analysis) yields a \emph{universal} profile $f(\zeta)$ (highlighted in gold) towards which the solution $h(x,t)$ evolves in the intermediate period after the IC is forgotten (but prior to leveling).}
\label{fig:introSS}
\end{figure}

Being governed by a parabolic (irreversible) equation, these currents `forget' their initial conditions after some time has elapsed; this is Barenblatt's concept of \emph{intermediate asymptotics} \cite{Barenblatt1972,Barenblatt1979}. Moreover, the PDE~\eqref{eq:nl_diff} can be reduced to an ordinary differential equation (ODE) through a self-similarity transformation. If the similarity variable can obtained by a scaling (dimensional) analysis, then the solution is termed a self-similar solution of the \emph{first} kind \cite[Ch.~3]{Barenblatt1979}. Specifically, the transformation is  $h(x,t) = \mathfrak{C}\, t^\beta f(\zeta)$ ($h$ in units\footnote{Throughout the chapter, we use SI units for all dimensional quantities.} of meters), where $\zeta = x/(\eta_N t^\delta)$ is the similarity variable (dimensionless), $f(\zeta)$ is the self-similar profile to be determined by solving an ODE, and $\mathfrak{C}$ and $\eta_N$ are dimensional consistency constants. The exponents $\beta$ and $\delta$ are obtained through scaling (dimensional analysis) of the governing PDE. As a representative example, consider the one-dimensional (1D) spreading of a fixed mass of fluid having an arbitrary `wavy' initial shape, as shown in figure~\ref{fig:introSS}. Suppose the fluid's shape $h(x,t)$ is governed by the linear diffusion equation $\partial h/\partial t = A \partial^2 h/\partial x^2$ (taking $A=1$ m\textsuperscript{2}/s in this example without loss of generality) subject to $(\partial h/\partial x)|_{x=\pm L}=0$. The initial condition (IC) is quickly `forgotten,' and the ultimate asymptotic state (here, flat) is achieved after passing through an intermediate asymptotic regime. It is straightforward to determine the self-similarity transformation: $\beta=-\delta=-1/2$ and $f(\zeta)=\mathrm{e}^{-\zeta^2}$ (see, e.g., \cite{Barenblatt1979} and the Appendix). Here, $\mathfrak{C}$ depends on the initial condition and $\eta_N=2$. The convergence of the rescaled $h(x,t)$ profiles towards $f(\zeta)$ can be clearly observed in figure~\ref{fig:introSS}(b). The IC is forgotten, and the profile converges onto the Gaussian intermediate asymptotic shape \cite{Barenblatt1979}. The profile $f(\zeta)$ is termed `universal' because it is independent of $h(x,0)$.

Having illustrated the notion of first-kind self-similarity as intermediate asymptotics, let us summarize its use in studying the gravitational  spreading of Newtonian viscous fluids in a variety of physical scenarios. For example, gravity currents arise in geophysical applications associated with flows through porous rocks \cite{Woods2015} such as in ground water extraction \cite{Bear1972}, during oil recovery \cite{Huppert2000,FELISA2018}, and during CO\textsubscript{2} sequestration \cite{Huppert2014}. In these examples, $h(x,t)$ represents an interface between two immiscible fluids in the limit of large Bond number (gravity dominates surface tension). There is now an extensive literature featuring a wealth of exact and approximate analytical self-similar solutions for gravity currents in porous media, e.g., \cite{Barenblatt1952,Huppert1995,Anderson2003,Lyle2005,Vella2006,Hesse2007,Anderson2010,DeLoubens2011,Ciriello2013,Huppert2013,zcs14} amongst many others. 

In this chapter, we focus on the propagation of non-Newtonian gravity currents, specifically ones for which the denser fluid obeys a \emph{power-law} rheology. This tractable model of non-Newtonian rheological response is also known as the Oswald--de Weale fluid \cite{BAH87}. In unidirectional flow, the power-law model simply dictates that fluid's viscosity depends upon a power of the velocity gradient.  \citet{DiFed2006} generalized Huppert's problem \cite{Huppert1982a} to power-law fluids, although Gratton et al~\cite{Gratton1999,Perazzo2003} had also considered some related problems. Even earlier, \citet{Kondic1996,Kondic1998} derived the governing equations for power-law fluids under confinement (i.e., in Hele-Shaw cells) using the lubrication approximation. These works have contributed to the use of a modified Darcy law to model the flow of non-Newtonian fluids in porous media using the analogy to flow in Hele-Shaw cells. \citet{aj_nonNewt1992} were perhaps the first to combine a Darcy law for a power-law fluid with the continuity equation to obtain a single PDE, of the kind studied herein, governing the gravity current's shape. Recently, \citet{Lauriola2018} highlighted the versatility of this approach by reviewing the existing literature and extending it to two-dimensional axisymmetric spreading in media with uniform porosity but variable permeability. All these flows are of interest because exact analytical self-similar solutions in closed form have been derived previously \cite{Gratton1999,Perazzo2005,DiFed2012,DiFed2017,Ciriello2016}. Specifically, the solution of \citet{Ciriello2016} will be used in \S\ref{sec:conv} below to verify the truncation error of the proposed numerical method.

For a self-similar solution to exist, both the governing PDE and its boundary conditions (BCs) must properly transform into an ODE in $\zeta$ with suitable BCs. A number of studies have specifically shown that the volume of fluid within the domain can be transient, varying as a power law in time, $\mathcal{V}(t) \propto t^\alpha$ ($\alpha \ge 0$), and a self-similar solution still exists  (see, e.g., \cite{Barenblatt1952,Lyle2005,Hesse2007,DiFed2012,zcs14,DiFed2017} and the references therein). However, the nonlinear ODE in $\zeta$ often cannot be integrated exactly in terms of known function, except for $\alpha=0$. In \S\ref{sec:eq} below, we discuss how a constraint of the form  $\mathcal{V}(t) \propto t^\alpha$ can be implemented numerically through flux BCs at the computational domain's ends. 

\begin{figure}[t]
    \centering
    \includegraphics[width=\textwidth,scale=0.95]{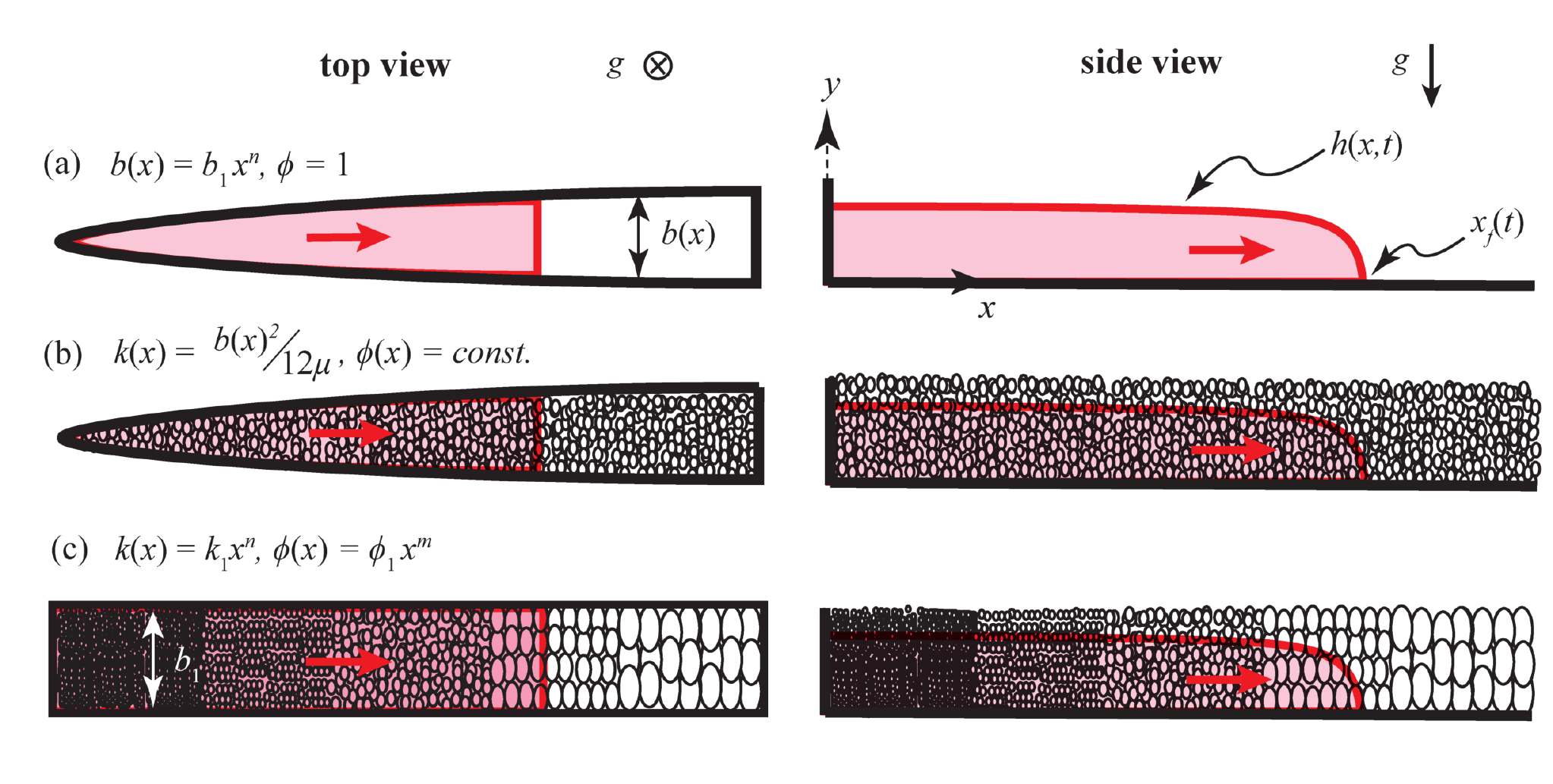}
    \caption{A summary of the gravity current flows and domains considered in this work. (a) Flow away from the origin in a completely porous ($\phi = 1$) HS cell of variable width given by $b(x) = b_1 x^n$ ($b_1 = const.$, $0\leq n<1$). (b) Flow in uniformly porous ($\phi = \phi_1 = const. \neq 1$) passage of variable width given by the same $b(x)$ as in (a). (c) Flow in a uniform-width slab (i.e., $b(x) = b_1 = const.$) with horizontally heterogeneous porosity and permeability given by $\phi (x) = \phi_1 x^m$ and $k(x) = k_1 x^n$,
    respectively. The effective permeability of the medium in (a) and (b) is set by the Hele-Shaw analogy via the width: $k(x) = [b(x)]^2/(12\mu)$. Figure reproduced and adapted with permission from [Zheng et al, Influence of heterogeneity on second-kind self-similar solutions for viscous gravity currents, J.\ Fluid Mech., vol.\ 747, p.\ 221] \textcopyright~Cambridge University Press 2014.}
    \label{fig:domains}
\end{figure}

With increasing complexity of the flow physics incorporated in the model, finding a self-similarity transformation may no longer be possible simply by  scaling (dimensional) arguments. \citet{Gratton1990} classified a number of such situations, including the so-called `focusing' flows involving fluid axisymmetrically flowing towards the origin on a flat planar surface. Further examples involving confined currents in channels with variable width, and/or in porous media whose permeability and porosity are functions of $x$, were proposed by \citet{zcs14}, as illustrated in  figure~\ref{fig:domains}. These gravity currents do enter a self-similar regime, even though a self-similar transformation cannot be obtained by scaling arguments alone. The exponents $\beta$ and $\delta$ in the transformation are unknown \textit{a priori}, hence this situation represents a self-similarity of the \emph{second} kind \cite[Ch.~4]{Barenblatt1979}. The governing equation can be transformed to an ODE, following which a nonlinear eigenvalue problem must be solved for $\beta$ and $\delta$ through a phase-plane analysis \cite{Gratton1990,Gratton1991}. Alternatively, experiments or numerical simulations are necessary to determine $\beta$ and $\delta$. For example, early numerical simulations were performed to this end by \citet{Diez1992}. However, a `pre-wetting film' ahead of the current's sharp wavefront ($x=x_f(t)$ where $h\big(x_f(t),t\big) = 0$) was required to avoid numerical instabilities. The scheme therein was also first-order accurate in time only. In this chapter, we propose a modern, high-order-accurate implicit numerical method for use in such problems.

Specifically, we develop and benchmark a strongly implicit and conservative numerical scheme for 1D nonlinear parabolic PDEs arising in the study of gravity currents. We show how the proposed scheme can be used to simulate (with high accuracy and at low computational expense) the spreading of 1D  \emph{non-Newtonian} viscous gravity currents in variable geometries (specifically, Hele-Shaw cells with widths varying as a power law in $x$). To this end, we build upon the work of \citet{zcs14}, which introduced this type of finite-difference scheme for simulating the spreading of a finite mass of Newtonian fluid in a variable-width Hele-Shaw cell. Owing to its accuracy and stability, this finite-difference scheme has been recently applied by \citet{koch_2018} to study hydraulic fracturing of low-permeability rock.

This chapter is organized as follows. In \S\ref{sec:prelim}, we briefly summarize existing models describing certain flows of viscous gravity currents. Then, we introduce a convenient general notation for such nonlinear parabolic PDEs. In \S\ref{sec:grid}, we introduce the 1D equispaced but staggered grid upon which the proposed finite-difference scheme is to be implemented. The derivation of the BCs for the PDE, from the mass conservation constraint, is discussed in \S\ref{sec:eq}. Then, we construct the nonlinear Crank--Nicolson scheme in \S\ref{sec:scheme_main} and discuss the discretized form of the nonlinear flux BCs in \S\ref{sec:BC}. Continuing, in \S\ref{sec:conv}, the scheme's accuracy is justified by comparing the numerical solution provided by the finite-difference scheme (up to a specified physical time) against an analytical solution obtained through a self-similar transformation of the PDE. Specifically, this approach involves three validation cases: (i) a symmetric (about $x=0$) lump of fixed fluid mass spreading in two directions (convergence is independent of BCs), (ii) a fixed fluid mass spreading away from the origin ($x=0$) (requires only no flux BCs), and (iii) a variable fluid mass injected at the origin spreading away from it (requires careful implementation of the nonlinear BCs). In all three cases, the scheme is shown to be capable of accurately computing the evolution of gravity current's shape. In \S\ref{sec:consv}, we analyze the scheme's conservation properties by verifying numerically that it respects the mass constraint $\mathcal{V}(t) \propto t^\alpha$. We consider two validation cases: (i) release of a fixed fluid mass ($\alpha = 0$), and (ii) fluid mass injection into the domain ($\alpha > 0$). In both cases, we specifically focus on the challenging case of a non-Newtonian (power-law) displacing fluid in a variable-width channel. As a benchmark, we use previously derived first-kind self-similar solutions from the literature, which are discussed in the Appendix.

\section{Preliminaries}
\label{sec:prelim}

In this section, we summarize the mathematical model for viscous gravity currents in a selected set of applications involving Newtonian and non-Newtonian fluids. We study their spreading in a fixed- or variable-width channel geometry (also known as a ``Hele-Shaw cell''), as well as flows in heterogeneous porous media with independently variable permeability and porosity. Our goal is to highlight the fact that all these models can be concisely summarized by a single nonlinear parabolic PDE supplemented with a set of nonlinear Neumann (flux) BCs.  

\subsection{Fluid Domain and Flow Characteristics}

The flow domain is assumed to be long and thin. For example, it can be a channel existing in the gap between two impermeable plates, i.e., a Hele-Shaw (HS) cell, which may or may not have variable transverse (to the flow) width as shown in figure~\ref{fig:domains}(a); or, it can be slab of uniform-thickness heterogeneously porous material, as shown in figure~\ref{fig:domains}(c). The viscous gravity current consists of one fluid displacing another immiscible fluid. Therefore, a sharp interface $y=h(x,t)$ separates the two fluids at all times. The present study considers the limit of negligible surface (interfacial) tension (compared to gravitational forces). The density difference $\Delta \rho$ between the two fluids is large compared to the density of the lighter fluid, and the denser fluid flows along the bottom of the cell, which is a horizontal impermeable surface. In doing so, the denser fluid displaces the lighter fluid out of its way. Here, the geometry is considered to be vertically unconfined so that the details of the flow of the upper (lighter) fluid can be neglected. 

We are interested in the evolution of the interface $h(x,t)$ between the two fluids. Owing to the vertically unconfined, long and thin geometry of the flow passage, the denser fluid has a slender profile (small aspect ratio), and the fluid flow can be described by \emph{lubrication theory}. The lubrication approximation also requires that viscous forces dominate inertial forces; this is the limit of small Reynolds number. In this regime of small Reynolds number but large Bond number, the flow is governed by a balance of viscous forces and gravity. Furthermore, the lubrication approximation allows for (at the leading order in the aspect ratio) the variation of quantities across the transverse direction, as well as the vertical velocities of the fluids to be neglected.  

As shown in figure~\ref{fig:domains}(a), for the flow in a HS cell, we allow the cell's width to vary as a power-law of the streamwise coordinate $x$, i.e., $b(x) = b_1 x^n$, where $n \geq 0$ is a dimensionless exponent, and $b_1>0$ is a dimensional consistency constant having units m$^{1-n}$. Since the cell has a variable width, it originates from a cell `origin,' which is always taken to be $x=0$ such that $b(0)=0$. As discussed in \cite{zcs14}, in such a flow geometry, the lubrication approximation may fail when $b(x)$ is an increasing function of $x$ i.e., $\mathrm{d} b/\mathrm{d} x = n b_1 x^{n-1} > 1$. In such quickly-widening cells, the transverse variations of properties become significant. We ensure the validity of the lubrication approximation, and models derived on the basis of it, by only considering $n<1$ such that $\mathrm{d} b/\mathrm{d}x$ remains a decreasing function of $x$.

The porosity can also be varied by filling the HS cell with beads of fixed diameter, as illustrated in figure~\ref{fig:domains}(b). We also consider a gravity current spreading horizontally in a porous slab of constant transverse width ($b(x) = b_1 = const.$) with heterogeneous porosity $\phi(x) = \phi_1 x^m$ and permeability $k(x) = k_1 x^n$, as shown in figure~\ref{fig:domains}(c). Here, $m,n\geq0$ are dimensionless exponents and $\phi_1, k_1>0$ are dimensional constants needed for consistency with the definitions of porosity and permeability,  respectively; specifically $\phi_1$ has units of units of m$^{-m}$, and $k_1$ has units of m$^{2-n}$. These variations are illustrated by the streamwise changes of  bead radii in figure~\ref{fig:domains}(c). Now, the point at which the porosity and permeability vanish is the origin of the cell. Another interesting case, that of a medium with vertically heterogeneous porosity, has been explored by \citet{Ciriello2016}. In this chapter, we limit our discussion to flow in a completely porous (i.e., unobstructed, $\phi=1$) HS cell of variable width as in figure~\ref{fig:domains}(a). However, the numerical scheme developed herein can readily treat any of these cases, taking the appropriate parameter definitions from  table~\ref{tb:eq_models} in \S\ref{sec:eq}.

We allow the denser fluid to be non-Newtonian. Specifically, it obeys the power-law rheology. In unidirectional flow, the one unique non-trivial shear stress component is given by $\tau = \mu (\dot \gamma) \dot \gamma$, where the dynamic viscosity $\mu$ depends on the shear rate $\dot \gamma$ as $\mu(\dot \gamma) = \mu_0 \dot \gamma^{r-1}$. Here, $\mu_0$ is the flow consistency index (units of Pa$\cdot$s$^{r}$), and $r$ ($>0$) is the fluid's rheological index. Fluids having $r<1$ are termed shear-thinning (e.g., blood), and fluids with $r>1$ are termed shear-thickening (e.g., dense particulate suspensions). In the special case $r=1$, the power-law model reduces to the Newtonian fluid. As stated above, the flow of the displaced fluid is immaterial to the dynamics of the gravity current, as long as the viscosity and density contrasts are large. This condition is satisfied, e.g., by assuming (for the purposes of this chapter) the displaced fluid is air.

Finally, the volume of the fluid in the cell itself may be either fixed (constant mass) or vary with time (injection). Consistent with the literature, we consider the instantaneous volume of fluid in the cell to increase as a power law in $t$: $\mathcal{V}(t) = \mathcal{V}_0 + \mathcal{V}_\mathrm{in} t^\alpha$, where $\mathcal{V}_0$ is the initial volume of fluid in the HS cell (measured in m\textsuperscript{3}), $\alpha \geq 0$ is a dimensionless exponent, and $\mathcal{V}_{in}$ is an injection pseudo-rate (in units m$^{3}$s$^{-\alpha}$), becoming precisely the injection rate for $\alpha=1$. Next, we discuss how this assumption leads to BCs for the physical problem and for the numerical scheme. 

\subsection{Governing Equation, Initial and Boundary Conditions}
\label{sec:eq}

The propagation of a viscous gravity current is described by a diffusion equation for the interface $h(x,t)$, which is the shape of profile of the denser fluid. The models are derived either from porous medium flow under Darcy's law and the Dupoit approximation \cite[Ch.~8]{Bear1972} or using lubrication theory with no-slip along the bottom of the cell and zero shear stress at the fluid--fluid interface \cite[Ch.~6-C]{L07}. The resulting velocity field is combined with a depth-averaged continuity equation to derive the nonlinear parabolic PDE for $h(x,t)$. We propose to summarize all gravity current propagation along horizontal surfaces through a single `thin-film' \cite{ODB97} equation:
\begin{equation}
    \frac{\partial h}{\partial t} = \frac{A}{x^p} \frac{\partial}{\partial x}\left(x^q \psi \frac{\partial h}{\partial x}\right).
\label{eq:nl_diff}
\end{equation}
According to \citet[Ch.~5]{Engl15}, eq.~\eqref{eq:nl_diff} can be classified as an `evolution equation.' 
The term in the parentheses on the right-hand side of eq.~\eqref{eq:nl_diff}, roughly, represents a fluid flux balanced by the change in height on the left-hand side. The multiplicative factor $A/x^p$ arises due to (i) geometric variations of the flow passage in the flow-wise direction, (ii) porosity variations in the flow-wise direction, or (iii) from the choice of coordinate system in the absence of (i) or (ii). Here, $A$ is dimensional constant depending on the flow geometry, the domain, and the fluid properties. Additionally, $p$ and $q$ are dimensionless exponents that depend on the flow geometry and fluid rheology. The quantity denoted by $\psi$ represents specifically the \emph{nonlinearity} in these PDEs. Thus, it is necessarily a function of $h$, and possibly $\partial h/\partial x$ for a non-Newtonian fluid (as in the third and fifth rows of table~\ref{tb:eq_models}).\footnote{Interestingly, an `$r$-Laplacian' PDE, similar to eq.~\eqref{eq:nl_diff} for a power-law fluid in a HS cell (third row of table~\ref{tb:eq_models}), arises during fluid--structure interaction between a power-law fluid and an enclosing slender elastic tube \cite{Boyko2017}. This PDE can also be tackled by the proposed finite-difference scheme.}

As stated in \S\ref{sec:intro}, several versions of eq.~\eqref{eq:nl_diff} will be explored herein, incorporating geometric variations, porosity variations, non-Newtonian behavior. The pertinent physical scenarios that will be tackled herein (using the proposed numerical scheme) are presented in table~\ref{tb:eq_models}, which lists expressions for $A$, $p$, $q$ and $\psi$. From a dimensional analysis of the PDE~\eqref{eq:nl_diff}, it follows that the constant $A$ must have units of m$^{1+p-q}\cdot$s$^{-1}$, as long as the nonlinearity $\psi$ has units of length (as is the case for all the models summarized in Table~\ref{tb:eq_models}). 
It is worth noting that in the case of 1D, linear diffusion ($p=q=0$ and $\psi = 1$), $A$ becomes the `diffusivity' in units of m$^2$/s.

The PDE~\eqref{eq:nl_diff} is solved on the finite space-time interval $(x,t) \in (\ell,L)\times(t_0,t_f]$. Here, $t_0$ and $t_f$ represent the initial and final times of the numerical simulation's run, respectively. An initial condition (IC) $h_0(x)$ is specified at $t=t_0$, so that $h(x,t_0)=h_0(x)$ is known. Meanwhile, $\ell$ is a small positive value (close to 0). Boundary conditions (BCs) are specified at $x=\ell$ and $x=L$. These involve some combination of $h$ and $\partial h/\partial x$. The reason for taking $x=\ell \ne 0$ becomes clear below.

Thus, let us now discuss such a suitable set of BCs. The BCs are based on the imposed mass conservation/growth constraint. Consider the case of a viscous gravity current in a porous slab with variable porosity $\phi(x) = \phi_1 x^m$, and transverse width $b_1=const.$ Then, the conservation of mass constraint (see \cite{zcs14}) takes the form
\begin{equation}
     \mathcal{V}(t) \equiv \int_\ell^L h(x,t) b_1 \phi(x) \, \mathrm{d}x = \mathcal{V}_0 + \mathcal{V}_\mathrm{in}t^\alpha,
    \label{eq:mass_conservn_porous}
\end{equation}
where $\alpha \geq 0$. In the parallel case of a HS cell with variable width $b(x) = b_1 x^n$ and porosity $\phi_1=const.$, which can either be set to unity or absorbed into $b(x)$ via $b_1$, the mass constraint becomes
\begin{equation}
    \mathcal{V}(t) \equiv \int_\ell^L h(x,t) b(x) \, \mathrm{d}x = \mathcal{V}_0 + \mathcal{V}_\mathrm{in}t^\alpha.
    \label{eq:mass_conservn}
\end{equation}
Taking a time derivative of eq.~\eqref{eq:mass_conservn} and employing eq.~\eqref{eq:nl_diff}, we obtain 
\begin{multline}
   \frac{\partial}{\partial t} \int_\ell^L h(x,t) b_1 x^n \, \mathrm{d}x
   =  \int_\ell^L \frac{\partial h}{\partial t} b_1 x^n \, \mathrm{d}x
   = \int_\ell^L b_1x^n \frac{A}{x^n}\frac{\partial}{\partial x}\left(x^q \psi \frac{\partial h}{\partial x}\right) \,\mathrm{d}x\\
   = Ab_1 \left.\left( x^q \psi \frac{\partial h}{\partial x} \right)\right|_{x=\ell}^{x=L}
   \;\stackrel{\text{by }\eqref{eq:mass_conservn}}{=}\; \frac{ \mathrm{d} (\mathcal{V}_\mathrm{in}t^\alpha)}{\mathrm{d} t} = \alpha \mathcal{V}_\mathrm{in}t^{\alpha-1}.
    \label{eq:BC_leibnitz}
\end{multline}
Here, $p=n$ in this case of interest, as described in table \ref{tb:eq_models}, and $Ab_1=const.$ Thus, we have obtained conditions relating $x^q\psi\partial h/\partial x$ at $x=\ell$ and $x=L$ to $\alpha\mathcal{V}_\mathrm{in}t^{\alpha-1}$. These conditions, if satisfied, automatically take into account the imposed volume constraint from eq.~\eqref{eq:mass_conservn}. The calculation starting with eq.~\eqref{eq:mass_conservn_porous} is omitted as it is identical, subject to proper choice of $p$.

For the case of propagation away from the cell's origin (i.e., any injection of mass must occur near $x=0$, specifically at $x=\ell$), to satisfy eq.~\eqref{eq:BC_leibnitz}, we can require that
\begin{subequations}\begin{align}
   \left.\left(x^q \psi \frac{\partial h}{\partial x}\right)\right|_{x = \ell} &= \begin{cases}
      - \frac{\alpha B}{A} t^{\alpha - 1}, &\quad \alpha > 0, \\[3pt]
      0, &\quad \alpha=0,
   \end{cases}\label{eq:L_bc}\displaybreak[3]\\[5pt]
   \left.\left(\psi \frac{\partial h}{\partial x}\right)\right|_{x=L} &= 0 \quad\Leftarrow\quad \left.\frac{\partial h}{\partial x}\right|_{x=L} = 0 ,
\end{align}\label{eq:bcs}\end{subequations}
where $B = \mathcal{V}_\mathrm{in}/b_1$. Recall, the case of $\alpha>0$ represents mass injection. Although eq.~\eqref{eq:mass_conservn} and eqs.~\eqref{eq:bcs} are equivalent, the imposition of the nonlinear BC in eq.~\eqref{eq:L_bc} must be approached with care. It should be clear that to impose a flux near the origin (at $x=0$), we need $\big(x^q \psi \partial h/\partial x\big)\big|_{x \to 0}$ to be finite. Then, $\psi \partial h/\partial x = \mathcal{O}(1/x^q)$ as $x\to0$. On the spatial domain $x \in (0,L)$, such an asymptotic behavior is possible for $p=q=0$. However, in a variable-width cell ($p,q \ne 0$), the local profile and slope as $x \to 0$ blow up if they are to satisfy $\psi \partial h/\partial x = \mathcal{O}(1/x^q)$ as $x\to0$. To avoid this uncomputable singularity issue, we defined the computational domain to be $x\in(\ell,L)$, where $\ell$ is `small' but $>0$. The BC from eq.~\eqref{eq:L_bc} at $x=\ell$ can then be re-written as
\begin{equation}
    \left.\left( \psi \frac{\partial h}{\partial x} \right)\right|_{x = \ell} = -\frac{\alpha B}{A\ell^{q}}t^{\alpha-1},\qquad \alpha > 0.
\end{equation}

It may also be of interest to consider the case of a gravity current released a finite distance away from the origin and then spreading towards $x=0$. In this case, an additional length scale arises in the problem: the initial distance of the current's edge from the origin, say $x_f(0)$. The existence of this extra length scale complicates the self-similarity analysis, leading to solutions of the second-kind \cite[Ch.~4]{Barenblatt1979}, as discussed in \S\ref{sec:intro}. However, the numerical scheme can handle this case just as well; in fact, it requires no special consideration, unlike spreading away from the origin. Now, we may simply take $\ell=0$ and consider spreading on the domain $(0,L)$ subject to the following BCs:
\begin{subequations}
\begin{align}
   \left.\left(x^q \psi \frac{\partial h}{\partial x}\right)\right|_{x\to 0} &= 0  \quad\Leftarrow\quad \left.\frac{\partial h}{\partial x}\right|_{x= 0} = 0, \label{eq:L_bc2}\\[5pt]
   \left.\left(\psi \frac{\partial h}{\partial x}\right)\right|_{x = L} &= \begin{cases}
      \frac{\alpha B}{A L^q} t^{\alpha - 1}, &\quad \alpha > 0, \\[3pt]
      0, &\quad \alpha=0,
   \end{cases}
\end{align}\label{eq:bcs2}\end{subequations}
which together allow us to satisfy eq.~\eqref{eq:BC_leibnitz} and, thus, eq.~\eqref{eq:mass_conservn} for all $t\in(t_0,t_f]$.

The most significant advantage of defining nonlinear flux BCs, such as those in eqs.~\eqref{eq:bcs} or \eqref{eq:bcs2}, is that a nonlinear nonlocal (integral) constraint, such as that in eq.~\eqref{eq:mass_conservn_porous} or \eqref{eq:mass_conservn}, no longer has to be applied onto the solution $h(x,t)$. Furthermore, if we start with compact initial conditions, i.e., there exists a nose location $x=x_f(t_0)$ such that $h\big(x_f(t_0),t_0\big)=0$, then the finite-speed of propagation property of the nonlinear PDE~\eqref{eq:nl_diff} \cite{GildKers2004,Vazquez2007} ensures that this nose $x_f(t)$ exists for all $t>t_0$ and $h\big(x_f(t),t\big)=0$ as well. The proposed fully-implicit scheme inherits this property of the PDE. Therefore, we can solve the PDE on the \emph{fixed} domain $x\in(\ell,L)$, without any difficulty, instead of attempting to rescale to a moving domain on which $x_f(t)$ is one of the endpoints with $h=0$ as the BC applied there. The latter approach proposed by \citet{Bonnecaze1992} (and used in more recent works \cite{Acton2001} as well) leads to a number of additional variable-coefficient terms arising in the PDE~\eqref{eq:nl_diff}, due to the non-Galilean transformation onto a shrinking/expanding domain. From a numerical methods point of view, having to discretize these additional terms is not generally desirable.

Having defined a suitable set of BCs, the last remaining piece of information required to close the statement of the mathematical problem at hand is the selection of pertinent initial conditions (ICs). For the case of the release of a finite fluid mass ($\alpha = 0$), an arbitrary polynomial IC may be selected, as long as it has zero slope at the origin ($x=0$), leading to satisfaction of the no-flux boundary condition~\eqref{eq:L_bc}. To this end, let the IC be given by
\begin{equation}
    h_0(x) = \begin{cases}
       a \left(\mathfrak{X}_0^c - x^c\right), &\quad x \leq \mathfrak{X}_0, \\[3pt]
      0, &\quad x>\mathfrak{X}_0,
   \end{cases}
   \label{eq:poly_IC}
\end{equation}
where $\mathfrak{X}_0$ is a `release-gate' location defining the initial position of the current's nose, i.e., $\mathfrak{X}_0=x_f(t_0)$ and $h\big(x_f(t_0),t_0\big)\equiv h_0(\mathfrak{X}_0)=0$. The constant $c > 1$ is an arbitrary dimensionless exponent. Finally, $a$ (units of m$^{1-c}$) is set by normalizing $h_0(x)$ such that the initial volume of fluid corresponds to the selected initial fluid volume, $\mathcal{V}_0$, via eq.~\eqref{eq:mass_conservn}.

The case of the release of a finite mass of fluid is particularly forgiving in how we set the IC, and its slope at $x=0$. In fact, we could even take $c=1$ in eq.~\eqref{eq:poly_IC} and the scheme will provide an initial flux of fluid at $t=t_0^+$, with $(\partial h/\partial x)_{x=0}=0$ thereafter. On the other hand, the case of mass injection ($\alpha > 0$) governed by the nonlinear BCs is not as forgiving. By virtue of the `point-source' mass injection at $x = \ell$, the slope at the origin rises sharply from the moment of mass injection. This very sharp rise has a tendency to introduce unphysical oscillations in the current profile when starting from the IC in eq.~\eqref{eq:poly_IC}. To avoid this, we must select a `better' IC, which has a shape more similar to the actual solution's singularity near $x=0$. Having tested a few different options, we found that an exponential function works well:
\begin{equation}
    h_0(x) = \begin{cases}
      a\left(-1 + b\mathrm{e}^{cx}\right), &\quad x \leq \mathfrak{X}_0, \\[3pt]
      0, &\quad x>\mathfrak{X}_0.
   \end{cases}
   \label{eq:expontial_IC}
\end{equation}
Here, $b$ (dimensionless) and $c$ (units of m$^{-1}$) are positive constants, $\mathfrak{X}_0 = \frac{1}{c}\ln{\frac{1}{b}}$ ensures that the IC has no negative values and a sharp wavefront, and $a$ (units of m) is set by normalizing $h_0(x)$ to the selected  intial volume $\mathcal{V}_0$ via eq.~\eqref{eq:mass_conservn}, as above.

Finally, it should be noted that the IC from eq.~\eqref{eq:poly_IC} is not used in the convergence studies for finite initial mass (\S\ref{sec:full_conv} and \S\ref{sec:haf_conv}). Rather, the IC is taken to be the exact self-similar solution of \citet{Ciriello2016} for a power-law fluid in a uniform-width ($n=0$) HS cell (see also the Appendix). The reasoning behind this  particular choice is further expounded upon in \S\ref{sec:conv}.

\begin{landscape}
\begin{center}
\begin{table}
\caption{Selected models of the propagation of viscous gravity currents herein simulated by a finite-difference scheme.}
\label{tb:eq_models}  
\begin{tabular}{p{5.2cm}p{5.2cm}p{1cm}p{2cm}p{2cm}}
\hline\noalign{\smallskip}
Case/Variable & $A$ [m$^{1-p+q}\cdot$s$^{-1}$] & $p$ [--] & $q$ [--] & $\psi$ [m] \\
\noalign{\smallskip}\svhline\noalign{\smallskip}
\begin{tabular}[c]{@{}l@{}}Newtonian fluid,\\fixed-width HS cell: $b(x) = b_1$. \\ {\scriptsize (see \citet{Huppert1995})} \end{tabular}  &  $\displaystyle\frac{\Delta\rho g b_1^2}{12\mu}$  & 0 & 0 & $h$\\ \noalign{\medskip}
\begin{tabular}[c]{@{}l@{}}Newtonian fluid,\\variable-width HS cell: $b(x) = b_1 x^n$. \\ {\scriptsize (see \citet{zcs14})} \end{tabular} & $\displaystyle\frac{\Delta\rho g b_1^2}{12\mu}$ & $n$ & $3n$ & $h$\\ \noalign{\medskip}
\begin{tabular}[c]{@{}l@{}}Power-law fluid: $\mu = \mu_0 \dot \gamma^{r-1}$,\\variable-width HS cell: $b(x) = b_1x^n$. \\ {\scriptsize (see \citet{DiFed2017,Longo2017})} \end{tabular} &$\displaystyle\left(\frac{r}{2r+1}\right)\left(\frac{\Delta\rho g}{\mu_0}\right)^{1/r}\left(\frac{b_1}{2}\right)^{(r+1)/r}$ & $n$ & $\displaystyle n\left(\frac{2r+1}{r}\right)$ & $\displaystyle h\left|\frac{\partial h}{\partial x}\right|^{(1-r)/{r}}$\\ \noalign{\medskip}
\begin{tabular}[c]{@{}l@{}}Newtonian fluid,\\ 2D porous medium, \\ variable porosity: $\phi(x) = \phi_1 x^m$, \\ variable permeability: $k(x) = k_1 x^n$.\\ {\scriptsize (see \citet{zcs14})} \end{tabular} & $\displaystyle\frac{\Delta \rho g k_1}{\mu \phi_1}$  & $m$ & $n$ & $h$\\ \noalign{\medskip}
\begin{tabular}[c]{@{}l@{}}Power-law fluid: $\mu = \mu_0 \dot \gamma^{r-1}$,\\ 2D porous medium, \\ variable porosity: $\phi(x) = \phi_1 x^m$, \\ variable permeability: $k(x) = k_1 x^n$.
\\ {\scriptsize (see \citet{Ciriello2016})} \end{tabular} & { $\displaystyle\begin{array}{lcl}2^{(3r+1)/2}\left(\frac{r}{3r+1}\right)^{1/r}\vphantom{\left(\frac{\Delta g}{\mu_0}\right)^{1/2}} \\ \qquad \times \left(\frac{k_1}{\phi_1}\right)^{(r+1)/2r}\left(\frac{\Delta \rho g}{\mu_0}\right)^{1/r}\end{array}$} & $m$ & $\frac{m(r-1) + n(r+1)}{2r}$ & $\displaystyle h\left|\frac{\partial h}{\partial x}\right|^{(1-r)/{r}}$\\
\noalign{\smallskip}\hline\noalign{\smallskip}
\end{tabular}
\newline
\end{table}
\end{center}
\end{landscape}

\section{The Numerical Scheme}
\label{sec:scheme}
The proposed numerical method is a finite-difference scheme using the Crank--Nicolson approach toward implicit time-stepping. Our presentation follows recent literature, specifically the construction in \cite[Appendix B]{zcs14}. The proposed scheme's truncation error is formally of second order in both space and time, and we expect the scheme to be unconditionally stable. Furthermore, the scheme is  conservative in the sense that it maintains the imposed time-dependency of the fluid volume with high accuracy via a specific set of nonlinear BCs. This section is devoted to discussing all these topics one by one.

\subsection{Notation: Grids, Time Steps, and Grid Functions}
\label{sec:grid}
\begin{figure}
    \centering
    \includegraphics[width=\textwidth,scale=0.75]{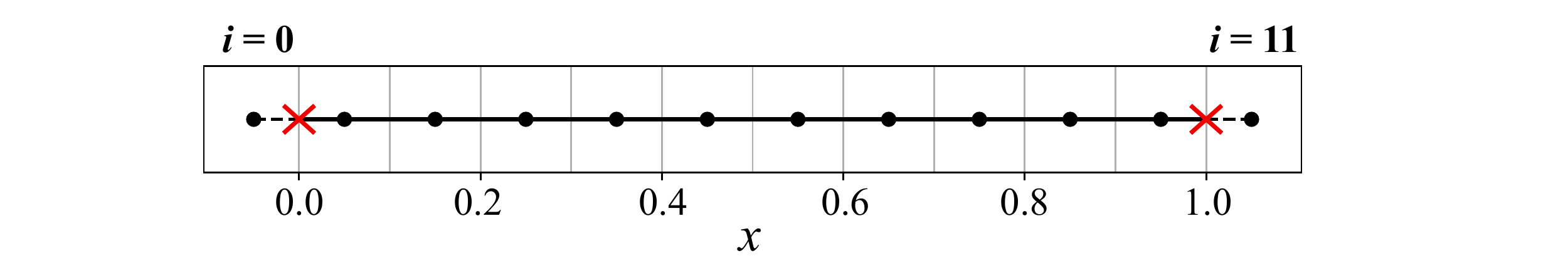}
    \caption{A sample twelve-node equispaced but staggered 1D grid. The grid nodes are staggered by half a grid step $\Delta x/2$ from the cell faces. The boundary conditions are implemented at the `real' domain boundaries (here marked by \textsf{x}). The two grid points  \emph{outside} the physical domain (i.e., $i = 0,11$ or  $x_0=-0.1$ and $x_{11}=1.1$ in this example) are used to implement the Neumann BCs, which require computing a derivative at the `real' domain boundaries (i.e., $i=1/2,21/2$ or  $x_{1/2} \equiv \ell=0$ and $x_{21/2} \equiv L=1$ in this example).}
    \label{fig:grid}
\end{figure}

The PDE \eqref{eq:nl_diff} is solved on an equispaced 1D grid of $N+1$ nodes with grid spacing $\Delta x = (L-\ell)/(N-1)$. The solution values are kept on a staggered grid of cell-centers, which are offset by $\Delta x/2$ with respect to the equispaced grid points. As a result, there is a node lying a half-grid-spacing beyond each domain boundary. It follows that the location of the $i$th grid point on the staggered grid is $x_i = \ell +  (i-1/2) \Delta x$, where $i = 0,1,2, \hdots, N$. A representative grid with 12 nodes is shown in figure \ref{fig:grid}. The use of a staggered grid affords additional stability to the scheme and allows us to evaluate derivatives with second-order accuracy via central differences, by default, using only two cell-centered values.

As stated in \S\ref{sec:eq}, the PDE  \eqref{eq:nl_diff} is solved over a time period $t \in (t_0, t_f]$, such that $t_f > t_0 \geq 0$, where both the initial time $t_0$ and the final time $t_f$ of the simulation are user defined. The scheme thus performs $M$ discrete time steps each of size $\Delta t = (t_f - t_0)/(M-1)$. The $n$th time step advances the solution to $t = t^n \equiv t_0 + n \Delta t$, where $n = 0, 1, \hdots, M-1$. Finally, we define the discrete analog (`grid function') to the continuous gravity current shape, which we actually solve for, as $h_i^n \approx h(x_i,t^n)$. 

\subsection{The Nonlinear Crank--Nicolson Scheme}
\label{sec:scheme_main}
Let us denote by $\mathcal{L}$ the continuous spatial operator acting on $h$ on the right-hand side of eq.~\eqref{eq:nl_diff}, i.e., 
\begin{equation}
\mathcal{L}[h] \equiv \frac{A}{x^p} \frac{\partial}{\partial x}\left(x^q \psi \frac{\partial h}{\partial x}\right).
\label{eq:Lcont}
\end{equation}
Since $\mathcal{L}$ is a second-order spatial operator and, thus, eq.~\eqref{eq:nl_diff} is a diffusion equation, we are inclined to implement a second-order-accurate time-stepping by the Crank--Nicolson scheme \cite{Crank1947}. The Crank--Nicolson scheme is fully implicit, which avoids the stringent restriction ($\Delta t \lesssim (\Delta x)^2$) suffered by explicit time discretizations of diffusion equations \cite[Ch.~6]{Strikwerda}. Then, the time-discrete version of eq.~\eqref{eq:nl_diff} is
\begin{equation}
\frac{h^{n+1}_i - h^n_{i}}{\Delta t} = \frac{1}{2}\big( \mathcal{L}_d\left[h_i^{n+1}\right] + \mathcal{L}_d\left[h_i^n\right] \big),
\label{eq:nl_diff_CNL}
\end{equation}
where $\mathcal{L}_d$ is the discrete analog to the continuous spatial operator $\mathcal{L}$ defined in eq.~\eqref{eq:Lcont}. Based on the approach of \citet{Christov2009}, the discrete spatial operator is constructed via flux-conservative central differencing using two cell-face values, while staggering the nonlinear terms:
\begin{subequations}
\begin{align}
\mathcal{L}_d\left[h_i^{n}\right] &= \frac{A}{x_i^p} \left[ \frac{\left(x_{i+1/2}^q\psi^{n+1/2}_{i+1/2}\right) S_{i+1/2}^n - \left(x_{i-1/2}^q \psi^{n+1/2}_{i-1/2}\right) S_{i-1/2}^n}{\Delta x} \right],\\
\mathcal{L}_d\left[h_i^{n+1}\right] &= \frac{A}{x_i^p} \left[ \frac{\left(x_{i+1/2}^q\psi^{n+1/2}_{i+1/2}\right) S_{i+1/2}^{n+1} - \left(x_{i-1/2}^q \psi^{n+1/2}_{i-1/2}\right) S_{i-1/2}^{n+1}}{\Delta x} \right],
\end{align}\label{eq:Ldiscrete}%
\end{subequations}
where $S \equiv \partial h/\partial x$ is the slope of the gravity current's shape.  Note that the nonlinear terms, denoted by $\psi$, have been evaluated the same way, i.e., at the mid-time-step $n+1/2$, for both $\mathcal{L}_d\left[h_i^{n}\right]$ and $\mathcal{L}_d\left[h_i^{n+1}\right]$.

Substituting eqs.~\eqref{eq:Ldiscrete} into eq.~\eqref{eq:nl_diff_CNL} results in a system of \emph{nonlinear} algebraic equations because $\psi$ is evaluated at mid-time-step $n+1/2$ and, thus, depends on both $h^{n}_i$ (known) and $h^{n+1}_i$ (unknown). This system must be solved for the vector $h^{n+1}_i$ ($i=0,\hdots,N$), i.e., the approximation to the gravity current's shape at the next time step. Solving a large set of nonlinear algebraic equations can be tedious and computationally expensive. A simple and robust approach to obtaining a solution of the nonlinear algebraic system is through fixed-point iterations, or `the method of internal iterations' \citep{Yanenko1971}. Specifically, we can iteratively compute approximations to  $h_i^{n+1}$, the grid function at the new time step, by replacing it in eq.~\eqref{eq:nl_diff_CNL} with $h_i^{n,k+1}$, where $h_i^{n,0}\equiv h_i^n$. Then, the proposed numerical scheme takes the form:
\begin{equation}
\begin{aligned}
\frac{h^{n,k+1}_i - h^n_{i}}{\Delta t} = &\frac{A}{2\Delta x}  \left[ {\frac{x_{i+1/2}^q}{x_i^p}}\psi^{n+1/2,k}_{i+1/2} S^{n,k+1}_{i+1/2} - {\frac{x_{i-1/2}^q}{x_i^p}}\psi^{n+1/2,k}_{i-1/2} S^{n,k+1}_{i-1/2} \right]\\
+ &\frac{A}{2\Delta x} \left[ {\frac{x_{i+1/2}^q}{x_i^p}}\psi^{n+1/2,k}_{i+1/2} S^{n}_{i+1/2} - {\frac{x_{i-1/2}^q}{x_i^p}}\psi^{n+1/2,k}_{i-1/2} S^{n}_{i-1/2} \right]. 
\end{aligned}
\label{eq:nontriag_scheme}
\end{equation}
The key idea in the method of internal iterations is to evaluate the nonlinear $\psi$ terms from information known at iteration $k$ and the previous time step $n$, while keeping the linear slopes $S$ from the next time step $n+1$ at iteration $k+1$. This manipulation linearizes the algebraic system, at the cost of requiring iteration over $k$. Upon convergence of the internal iterations, $h_i^{n+1}$ is simply the last iterate $h_i^{n,k+1}$. Before we can further discuss the iterations themselves or their convergence, we must define our discrete approximations for $\psi$ and $S$.

The operator $\mathcal{L}_d$ is essentially a second derivative, so we take inspiration from the standard way of constructing the three-point central finite-difference formula for the second derivative \cite{Strikwerda}. Therefore, $S_{i \pm 1/2}$ can be discretized using a two-point central-difference approximation on the staggered grid. For example, at any time step: 
\begin{equation}
S_{i+1/2} \equiv \left(\frac{\partial{h}}{\partial{x}}\right)_{x=x_{i+1/2}} \approx \frac{h_{i+1} - h_{i}}{\Delta x}.
\label{eq:dhdx_discr}
\end{equation}

\begin{figure}[t]
    \centering
    \includegraphics[width=0.75\textwidth]{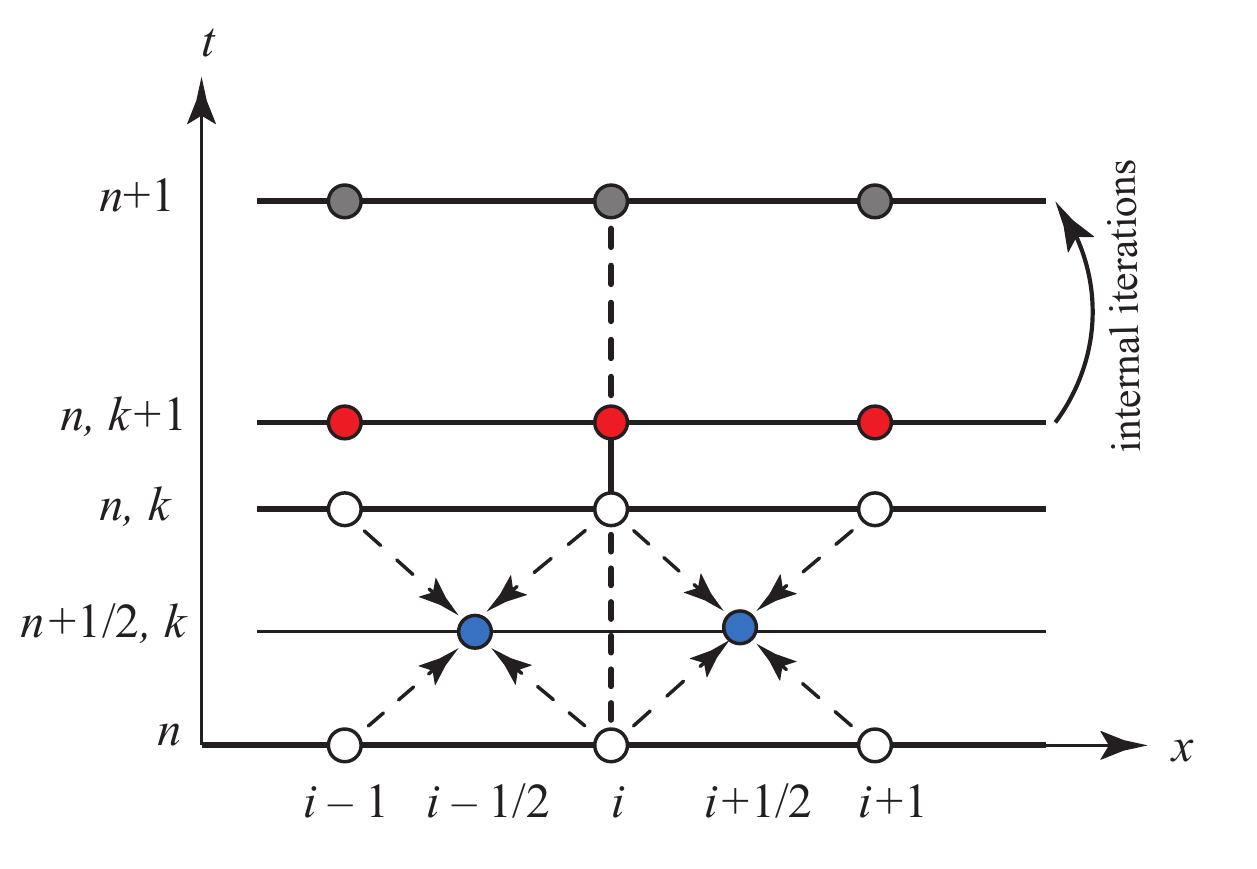}
    \caption{Representative stencil of the proposed scheme. After performing $k$ internal iterations, the nonlinear terms $\psi_{i \pm 1/2}$ are computed at the intermediate stage `$n+1/2,k$' (highlighted in blue) from the known quantities $h_i^n$ and $h_i^{n,k}$. The unknown quantity $h_i^{n,k+1}$ at the next internal iteration, stage `$n,k+1$' (highlighted in red), is found by solving the linear system in eq.~\eqref{eq:fin_diff}. The process continues until the convergence criterion in eq.~\eqref{eq:internal_conv} is met, yielding the (initially unknown) solution at $t=t^{n+1}$.}
    \label{fig:stencil}
\end{figure}

Next, following \cite{zcs14,Christov2002}, we evaluate $\psi$ at $x_{i\pm1/2}$ by averaging the known values at $x_i$ and $x_{i+1}$ or $x_i$ and $x_{i-1}$, respectively. Likewise, to approximate $\psi^{n+1/2}$, we average the known values: $\psi^n$ at $t^n$ and $\psi^{n,k}$ at the previous internal iteration. In other words, our approximation of the nonlinear terms is
\begin{subequations}\begin{align}
{\psi}^{n+1/2,k}_{i+1/2} &= \frac{1}{2}\Bigg[\underbrace{\frac{1}{2}\left({\psi}^{n,k}_{i+1} + {\psi}^{n,k}_{i}\right)}_{={\psi}^{n,k}_{i+1/2}} + \underbrace{\frac{1}{2}\left( {\psi}^{n}_{i+1} + {\psi}^{n}_{i}\right)}_{={\psi}^{n}_{i+1/2}}\Bigg], \label{eq:psi_outa}\\
{\psi}^{n+1/2,k}_{i-1/2} &= \frac{1}{2}\Bigg[\underbrace{\frac{1}{2}\left({\psi}^{n,k}_{i} + {\psi}^{n,k}_{i-1}\right)}_{={\psi}^{n,k}_{i-1/2}} + \underbrace{\frac{1}{2}\left( {\psi}^{n}_{i} + {\psi}^{n}_{i-1}\right)}_{={\psi}^{n}_{i-1/2}}\Bigg].
\label{eq:psi_outb}
\end{align}
\label{eq:psi}%
\end{subequations}
Equations~\eqref{eq:psi} afford improved stability for nonlinear PDEs, while preserving the conservative nature of the scheme (as will be shown in \S\ref{sec:consv}), as discussed by \citet{VonR1975} who credits the idea of averaging nonlinear terms across time stages and staggered grid points to the seminal work of \citet{Douglas1959,Douglas1962}. The scheme thus described is depicted by the stencil diagram in Figure~\ref{fig:stencil}.

Here, it is worthwhile noting that, the classical Crank--Nicolson \cite{Crank1947} scheme is only \emph{provably} unconditionally stable \cite{Strikwerda} when applied to a \emph{linear} diffusion equation. It was suggested by \citet{Christov2009} that the current approach provides additional stability to this \emph{nonlinear} scheme for large time steps. But, since our problem is nonlinear, some care should be taken in evaluating how large of a time step could be taken. Nevertheless, it is still expected that the largest stable $\Delta t$ will be independent of $\Delta x$.

A complication arising in the present context is that we focus on the case of a power-law non-Newtonian viscous gravity current spreading in a variable-width cell. As a result, recalling table \ref{tb:eq_models}, this model features $\partial h/\partial x$ in $\psi$, \emph{unlike} the Newtonian case. While the temporal accuracy of the scheme is ensured through the robust implementation of the nonlinear Crank--Nicolson time-stepping, the spatial accuracy is  contingent upon the discretization of $\partial h/\partial x$ in $\psi$. A further consequence is that, once we discretize $\partial h/ \partial x$, the discretization of $\psi$ becomes \emph{nonlocal} (i.e., it requires information beyond the $i$th grid point). Nevertheless, the overall scheme still only requires a three-point stencil for $\mathcal{L}_d$. In particular, for interior grid points, we use a central-difference formula, giving rise to the expression (at any time step):
\begin{equation}
{\psi}_{i} \equiv \left[ h \left|\frac{\partial h}{\partial x}\right|^{(1-r)/{r}} \right]_{x=x_i} \approx h_{i}\left|\frac{h_{i+1} - h_{i-1}}{2\Delta x}\right|^{(1-r)/{r}}.
\label{eq:psi_disc}
\end{equation}
This choice of approximation ensures second-order accuracy at all interior grid nodes. However, at the second ($i=1$) and the penultimate ($i=N-1$) nodes, the second-order accurate approximation to $\partial h/\partial x$ in $\psi_{i\pm1/2}$ as defined in eqs.~\eqref{eq:psi} requires the unknown values $h_{-1}$ and $h_{N+1}$, respectively. To resolve this difficulty, we use  `biased' (backward or forward) three-point difference approximations:
\begin{subequations}\begin{align}
{\psi}_{0} &\approx h_{0}\left|\frac{-3h_{0} +4h_{1} - h_{2}}{2\Delta x}\right|^{(1-r)/{r}},\\
{\psi}_{N} &\approx h_{N}\left|\frac{3h_{N} -4h_{N-1} + h_{N-2}}{2\Delta x}\right|^{(1-r)/{r}}.
\end{align}\label{eq:psi_boundary}\end{subequations}

Finally, substituting the discretization for $S$ from eq.~\eqref{eq:dhdx_discr} into eq.~\eqref{eq:nontriag_scheme}, it is possible to re-arrange the scheme into a tridiagonal matrix equation:
\begin{multline}
\underbrace{\left[-\frac{A\Delta t}{2(\Delta x)^2} {\frac{x_{i-1/2}^q}{x_i^p}}\psi_{i-1/2}^{n+1/2,k}\right]}_{\text{matrix subdiagonal coefficient}} h^{n,k+1}_{i-1}\\ 
 + \underbrace{\left[1 + \frac{A\Delta t}{2(\Delta x)^2}\left({\frac{x_{i+1/2}^q}{x_i^p}}\psi^{n+1/2,k}_{i+1/2} + {\frac{x_{i-1/2}^q}{x_i^p}}\psi^{n+1/2,k}_{i-1/2}\right)\right]}_{\text{matrix diagonal coefficient}} h^{n,k+1}_i\\
+ \underbrace{\left[ - \frac{A\Delta t}{2(\Delta x)^2} {\frac{x_{i+1/2}^q}{x_i^p}}\psi_{i+1/2}^{n+1/2,k} \right]}_{\text{matrix superdiagonal coefficient}} h^{n,k+1}_{i+1}\displaybreak[3]\\
= h^n_{i} +  \frac{A\Delta t}{2(\Delta x)^2}\left[ {\frac{x_{i+1/2}^q}{x_i^p}}\psi^{n+1/2,k}_{i+1/2} (h^{n}_{i+1} - h^{n}_i) -  {\frac{x_{i-1/2}^q}{x_i^p}}\psi^{n+1/2,k}_{i-1/2}(h^{n}_{i} - h^{n}_{i-1}) \right]
\label{eq:fin_diff}
\end{multline} 
for the interior grid points $i=1,\hdots,N-1$. In eq.~\eqref{eq:fin_diff}, the right-hand side and the variable coefficients in brackets on the left-hand side are both known, based on $h_i^{n,k}$, at any given internal iteration $k$. Then, each internal iteration involves the inversion of a tridiagonal matrix to solve for the grid function $h_i^{n,k+1}$. The inversion of this tridiagonal matrix can be performed efficiently with, e.g., `backslash' in {\sc Matlab}. Subsequently, the coefficient matrix must be recalculated for each internal iteration because of the dependency of $\psi_{i\pm1/2}^{n+1/2,k}$ on $h_i^{n,k}$ arising from eqs.~\eqref{eq:psi}, \eqref{eq:psi_disc} and \eqref{eq:psi_boundary}

The iterations in eq.~\eqref{eq:fin_diff} are initialized with $h^{n,0}_i = h^n_i$ ($i=0,\hdots,N$) and continue until an iteration $k+1=K$ is reached at which a $10^{-8}$ relative error tolerance is met. Specifically, 
\begin{equation}
\max_{0\le i\le N}\left|h^{n,K}_i-h^{n,K-1}_i\right| < 10^{-8} \max_{0\le i\le N} \left|h^{n,K-1}_i\right|.
\label{eq:internal_conv}
\end{equation} 
Only a small number (typically, less than a dozen) of internal iterations are required at each time step, making the scheme quite efficient overall.

A detail remains, however. The algebraic system defined in eq.~\eqref{eq:fin_diff} applies to all \emph{interior} nodes, i.e., $i =  1,\hdots,N-1$.  To complete the system, we must define rows $i=0$ and $i=N$, which arise from the discretization of the nonlinear BCs, which comes in \S\ref{sec:BC}. Upon completing the latter task successfully, $h^{n,K}_i$ becomes the grid function at the next time step $h^{n+1}_i$ upon the completion of the internal iterations, and the time stepping proceeds.

\subsection{The Special Case of Linear Diffusion}
A noteworthy special case of the proposed finite-difference scheme arises from setting the dimensionless exponents $p=q=0$ (i.e., no spatial variation of the diffusivity) and $\psi = 1$ (linear diffusion). Then, eq.~\eqref{eq:fin_diff} can be simplified and rearranged in the form ($i=1,\hdots,N-1$):
\begin{equation}
    \left[1 + \frac{A \Delta t}{(\Delta x)^2}\right]h_{i}^{n+1}
    = \frac{A \Delta t}{(\Delta x)^2}\left(h_{i-1}^{n+1} + h_{i+1}^{n+1} + h_{i-1}^{n} + h_{i+1}^{n}\right) + \left[ 1 + \frac{A \Delta t}{(\Delta x)^2}\right]h_i^n.
\label{eq:CN_reduction}
\end{equation}

If the grid function $h_i^n \approx h(x_i,t^n)$ represents the temperature field along a 1D rigid conductor situated on $x\in[\ell,L]$, eq.~\eqref{eq:CN_reduction} is then the original second-order (in space and time) numerical scheme proposed by \citet{Crank1947} to solve a linear (thermal) diffusion equation \cite[\S6.3]{Strikwerda}. As such, this simplification helps illustrate the mathematical roots of the current scheme, and how we have generalized the classical work.

\subsection{Implementation of the Nonlinear Boundary Conditions}
\label{sec:BC}
As discussed in \S\ref{sec:eq}, the boundary conditions are a manifestation of the global mass conservation constraint, eq.~\eqref{eq:mass_conservn_porous} or \eqref{eq:mass_conservn}, imposed on eq.~\eqref{eq:nl_diff}. The BCs described in eqs.~\eqref{eq:bcs} and \eqref{eq:bcs2} are defined at the `real' boundaries of the domain, i.e., at $x=\ell$ and $x=L$. The numerical scheme is implemented over a staggered grid. This allows for derivatives at $x=\ell$ and $x=L$ to be conveniently approximated using central difference formulas using two nearby staggered grid points. In this manner, the BC discretization maintains the scheme's second order accuracy in space and time. Accordingly, for the case of a current spreading away from the cell's origin, eqs.~\eqref{eq:bcs} are discretized in a `fully-implicit' sense (to further endow numerical stability and accuracy to the scheme \cite{Christov2002}) as follows:
\begin{align}
   \psi_{1/2}^{n+1/2,k} \frac{1}{\Delta x}\left(h^{n,k+1}_1 - h^{n,k+1}_0\right) &= \left\{
    \begin{array}{ll}
      - \frac{\alpha B}{A \ell^q} t^{\alpha - 1}, &\quad \alpha > 0, \\[5pt]
      0, &\quad \alpha=0,
   \end{array} \right. \\[3pt]
   \frac{1}{\Delta x}\left(h^{n,k+1}_{N} - h^{n,k+1}_{N-1}\right) &= 0.
\end{align}
Within the internal iterations, however, $\psi_{1/2}^{n+1/2,k}$ is known independently of $h^{n,k+1}_1$ and $h^{n,k+1}_0$. Hence, we can express the first ($i=0$) and last ($i=N$) equations, which defined the respective rows in the tridiagonal matrix stemming from eq.~\eqref{eq:fin_diff}, as
\begin{subequations}\begin{align}
    h^{n,k+1}_1 - h^{n,k+1}_0 &= \left\{
    \begin{array}{ll}
      \displaystyle-\frac{4 \alpha B t^{\alpha-1}\Delta x}{A \ell^q (\psi_0^n + \psi_1^n + \psi_0^{n,k} + \psi_1^{n,k})}, &\quad\alpha > 0 , \\[5pt]
      0, &\quad\alpha=0,
   \end{array} \right. \\[3pt]
   h^{n,k+1}_{N} - h^{n,k+1}_{N-1} &= 0.
   \label{eq:num_bc_RHS}%
 \end{align}\label{eq:num_bc}\end{subequations}%

Similarly, we can derive the discretized BCs for spreading towards the origin, upon its release a finite distance away from the origin, from eqs.~\eqref{eq:bcs2}. Then, the first ($i=0$) and last ($i = N$) equations, which defined the respective rows in the tridiagonal matrix, as
\begin{subequations}\begin{align}
    h^{n,k+1}_{1}-h^{n,k+1}_{0} &= 0, \label{eq:num_bc2_LHS}\\
    h^{n,k+1}_{N} - h^{n,k+1}_{N-1} &= \left\{
    \begin{array}{ll}
      \displaystyle\frac{4 \alpha B t^{\alpha-1}\Delta x}{A L^q (\psi_{N-1}^n + \psi_{N-2}^n + \psi_{N-1}^{n,k} + \psi_{N-2}^{n,k})}, &\quad\alpha > 0 , \\[5pt]
      0, &\quad\alpha=0.
   \end{array} \right.
\end{align}\label{eq:num_bc2}\end{subequations}%

\section{Convergence and Conservation Properties of the Scheme}
\label{sec:prop}

At this point, the numerical scheme and boundary conditions defined in eqs.~\eqref{eq:fin_diff} and \eqref{eq:num_bc} or \eqref{eq:num_bc2} form a complete description of the numerical solution to the parabolic PDE from eq.~\eqref{eq:nl_diff}, for a gravity current propagating away from the origin. We have claimed that the finite-difference scheme is conservative (i.e., it accurately maintains the imposed time-dependency of the fluid volume set by eq.~\eqref{eq:mass_conservn}) and has second-order convergence. These aspects of the scheme will be substantiated in \S\ref{sec:conv} and \S\ref{sec:consv}, respectively. The computational domain's dimensions, which are set by $L$ and $b_1$, and the properties of fluid being simulated are summarized in table~\ref{tb:sim_para}. For definiteness, in this chapter we select the fluid properties to be those of a 95\% glycerol-water mixture in air at 20\textdegree C (see \cite{Cheng_vis_2008,VolkAndreas2018}).

\subsection{Estimated Order of Convergence}
\label{sec:conv}

First, we seek to justify the formal accuracy (order of convergence) of the proposed scheme through carefully chosen numerical examples. To do so, we pursue a series of benchmarks that are  successively `more complicated' (from a numerical perspective). First, we simulate the case of a centrally released fixed mass of fluid propagating in two directions (\S\ref{sec:full_conv}). Second, we simulate the unidirectional spreading of a fixed mass of fluid (\S\ref{sec:haf_conv}). Last, we simulate the unidirectional spreading of a variable fluid mass (\S\ref{sec:haf_conv_inject}) by taking into account injection of fluid at the boundary.

\begin{table}[t]
\caption{Summary of the simulation parameters used in convergence and conservation studies. The fluid was assumed to be a 95\% glycerol-water mixture at 20\textdegree C. The width exponent $n$ and fluid's rheological index $r$ were varied on a case-by-case basis to simulate different physical scenarios.}
\centering
\begin{tabular}{p{5cm}p{2cm}p{2cm}}
\noalign{\smallskip}\hline
\noalign{\smallskip}
Parameter & Value & Units\\
\noalign{\smallskip}\svhline\noalign{\smallskip}
Channel length, $L$             & 0.75    & m\\ \noalign{\smallskip}
Width coefficient, $b_1$    & 0.017390 & m$^{1-n}$ \\ \noalign{\smallskip} \hline \noalign{\smallskip}
Total released mass, $w$        & 0.31550  & kg\\ \noalign{\smallskip}
Density difference, $\Delta\rho$      & 1250.8  & kg/m\textsuperscript{3}\\ \noalign{\smallskip}
Consistency index ($r\ne1$)\\ or dynamic viscosity ($r=1$), $\mu_0$  & 0.62119 & Pa$\cdot$s$^r$ \\ \noalign{\smallskip} \hline
\end{tabular}
\label{tb:sim_para}
\end{table}

In each of these three cases, there is a need for a reliable benchmark solution against which the numerical solutions on successively refined spatial grids can be compared. For the case of the release of a fixed mass  of fluid, an exact self-similar solution is provided by \citet{Ciriello2016}. Specifically the solution is for a power-law fluid in uniform HS cell ($n=0$). The derivation of the self-similar solution is briefly discussed in the Appendix. We use this solution as the benchmark. As mentioned in \S\ref{sec:intro}, parabolic equations `forget' their IC and the solution becomes self-similar after some time. However, for a general PDE, it is difficult (if not impossible) to estimate how long this process takes. Therefore, to ensure a proper benchmark against the exact self-similar solution, we start the simulation with the exact self-similar solution evaluated at some non-zero initial time ($t_0>0$). Then, we let the current propagate up to a final time $t_f$, with the expectation that the current will remain in the self-similar regime for all $t\in[t_0,t_f]$. Comparing the final numerical profile with the exact self-similar solution at $t=t_f$ then allows for a proper benchmark.

To quantify the error between a numerical solution $h_\text{num}$ and a benchmark $h_\text{exact}$ solution at $t=t_f$, we use three standard function-space norms \cite{Evans2010}:
\begin{subequations}\begin{align}
\|h_\text{num}(x, t_f) - h_\text{exact}(x, t_f)\|_{L^\infty} &= \max_{x \in [\ell,L]} \left|h_\text{num}(x, t_f) - h_\text{exact}(x, t_f)\right|,\label{eq:Linf_cont}\\ 
\|h_\text{num}(x, t_f) - h_\text{exact}(x, t_f)\|_{L^1} &= \int_{\ell}^{L} \left|h_\text{num}(x, t_f) - h_\text{exact}(x, t_f)\right|\,\mathrm{d}x,\label{eq:L1_cont}\\
\|h_\text{num}(x, t_f) - h_\text{exact}(x, t_f)\|_{L^2} &= \sqrt{\int_{\ell}^{L} \left|h_\text{num}(x, t_f) - h_\text{exact}(x, t_f)\right|^2\,\mathrm{d}x}.\label{eq:L2_cont}
\end{align}\label{eq:Norms_cont}\end{subequations}
Using a second-order trapezoidal rule for the integrals, the definitions in eqs.~\eqref{eq:Norms_cont} can be expressed in terms of the grid functions to define the `errors':
\begin{subequations}\begin{align}
L^\infty_{\mathrm{error}} &\equiv \max_{0\le i\le N} \left|h_i^M - h_{\mathrm{exact}}(x_i,t_f)\right|,\label{eq:Linf}\\ 
\begin{split}
L^1_{\mathrm{error}} &\equiv \Delta x \left\{ \frac{1}{2}\Big[\left|h_0^M - h_{\mathrm{exact}}(x_0,t_f)\right| + \left|h_N^M - h_{\mathrm{exact}}(x_N,t_f)\right|\Big] \right.\\ &\hspace{3cm} + \sum_{i=1}^{N-1}\left. \left|h_i^M - h_{\mathrm{exact}}(x_i,t_f)\right| \vphantom{\frac{1}{2}}\right\},
\end{split}\label{eq:L1}\\ 
\begin{split}L^2_{\mathrm{error}} &\equiv \left[ 
\Delta x \left\{ \frac{1}{2}\left[\left|h_0^M - h_{\mathrm{exact}}(x_0,t_f)\right|^2 + \left|h_N^M - h_{\mathrm{exact}}(x_N,t_f)\right|^2\right] \right. \right.\\ & \hspace{3cm} + \sum_{i=1}^{N-1}\left.\left. \left|h_i^M - h_{\mathrm{exact}}(x_i,t_f)\right|^2 \vphantom{\frac{1}{2}}\right\}\right]^{1/2},
\end{split}\label{L2}
\end{align}\end{subequations}
where $M$ is the time step at which $t^M = t_f$.  

Since the solution actually has a corner (derivative discontinuity) at the nose (wavefront) $x_f(t)$ such that $h\big(x_f(t),t\big)=0$, the propagating gravity current is in fact only a \emph{weak} solution to the PDE \cite{Evans2010}. Therefore, the $L^\infty$ norm is not a good one to measure the error, as we do not expect the solution to `live' in this function space. Nevertheless, our  numerical results show convergence in the $L^\infty$ norm. The natural functional space for solutions of eq.~\eqref{eq:nl_diff} is the space of integrable functions, i.e., $L^1$. Indeed, we observe excellent second-order convergence in this norm. For completeness, the $L^2$ norm (commonly the function-space setting for parabolic equations \cite[Ch.~7]{Evans2010}) is considered as well. While we observe convergence close to second order in this norm as well, it is clearly not the `natural' one for these problems either.

For our estimated-order-of-convergence study, $\Delta x$ is successively halved on a domain of fixed length, such that on the $c$th iteration of the refinement, the grid spacing is $\Delta x_c = \Delta x_0/2^{c-1}$, where $\Delta x_0$ is the initial grid spacing. Doing so ensures a set of common grid points (corresponding to the same physical locations) between successively refined grids. In all studies in this section, we begin with a grid with $N = 101$ nodes, and it becomes the coarsest grid for the refinement study. Given the (formally) unconditionally stable nature of the scheme, we take $\Delta t_c = 2\Delta x_c$ for the refinement studies without loss of generality. From a computational standpoint, it is desirable that time step and grid spacing are of the same order of magnitude in the estimated-order-of-convergence study.

\subsubsection{Central Release of a Fixed Fluid Mass (No Boundary Effects)}
\label{sec:full_conv}

Consider a symmetric domain $x\in[-L,+L]$. Then suppose that a fixed mass of fluid (i.e., $\alpha = 0$ in the volume constraint in eq.~\eqref{eq:mass_conservn}) is released with an initial shape that is symmetric about $x=0$. The final simulation time $t_f$ is such that the gravity current does not reach $x=\pm L$ for $t \le t_f$. Since the fluid mass is constant and the BCs are imposed at $x=\pm L$ (where $h=0$ initially and remains so for all $t\le t_f$, by construction), their discretization simply reduces to the trivial cases, i.e., eqs.~\eqref{eq:num_bc2_LHS} and \eqref{eq:num_bc_RHS}. Thus, the BCs for this study are simply linear Neumann (i.e., no flux or homogeneous) BCs, and they do not influence the order of convergence of the overall scheme. Therefore, this study allows us to verify that our approach to the treatment of the nonlinearity $\psi$, and its weighted averages appearing in the spatially discretized operator $\mathcal{L}_d$ in eq.~\eqref{eq:Ldiscrete}, deliver the desired second-order of accuracy in space. Coupled with the Crank--Nicolson time-stepping's second-order accuracy in time, we thus expect second order of convergence in this refinement study.

As stated above, we take the exact self-similar solution to eq.~\eqref{eq:nl_diff} provided by \citet{Ciriello2016} (and discussed in the Appendix) evaluated at $t=t_0$ and mirrored about $x=0$ as the IC. Upon evolving this IC numerically up to $t=t_f$, we compare the numerical profile to the same exact solution now evaluated at $t=t_f$. Hence, in accordance with the assumptions required to obtain this exact solution in \cite{Ciriello2016}, we limit this first convergence study to a uniform-width HS cell, i.e., $n=0$.

\begin{figure}[t]
\centering
\subfloat[][Newtonian fluid ($r=1$).]{\includegraphics[width=0.5\textwidth]{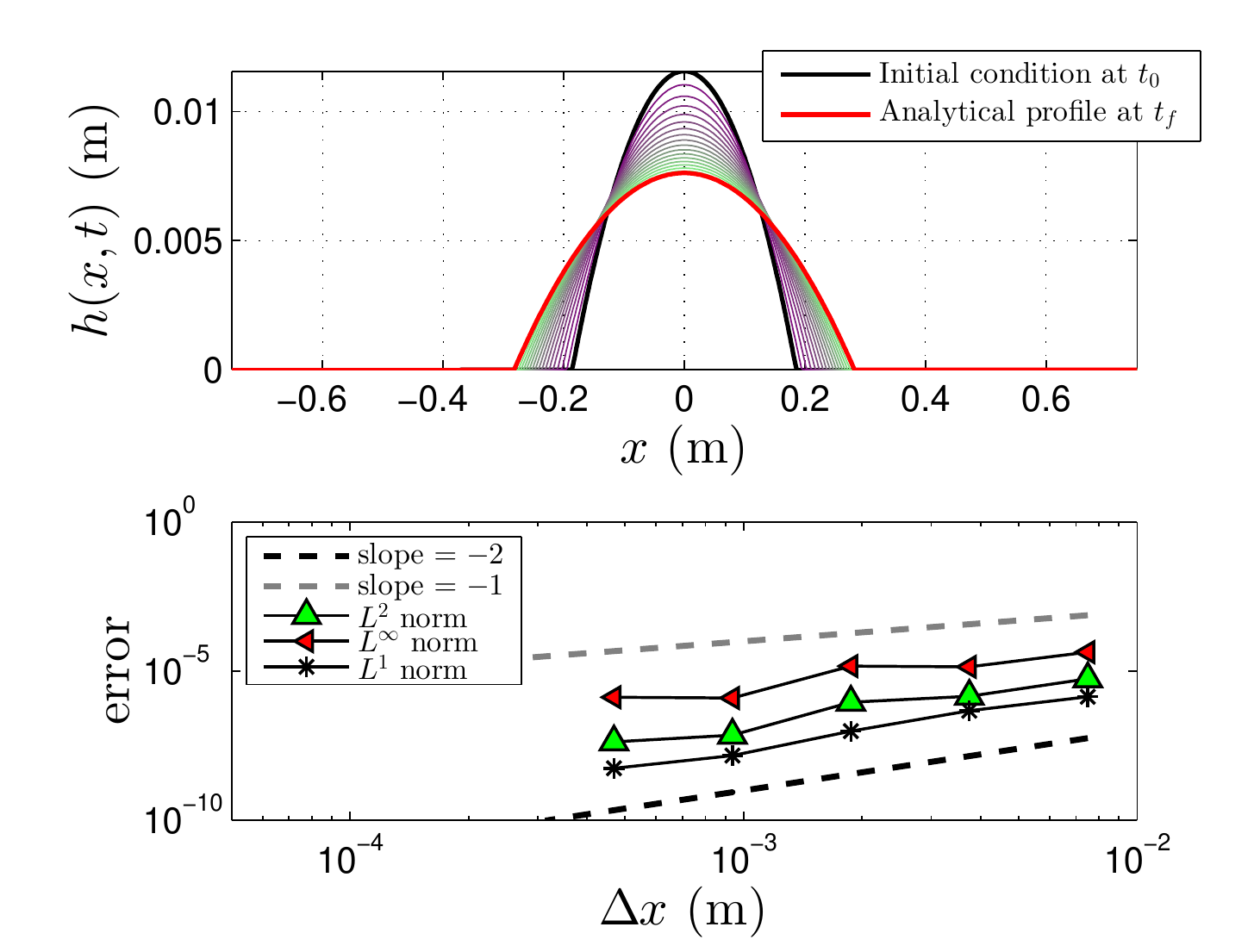}\label{fig:centnewt}}\\
\subfloat[][Shear-thinning fluid ($r=0.7$).]{\includegraphics[width=0.5\textwidth]{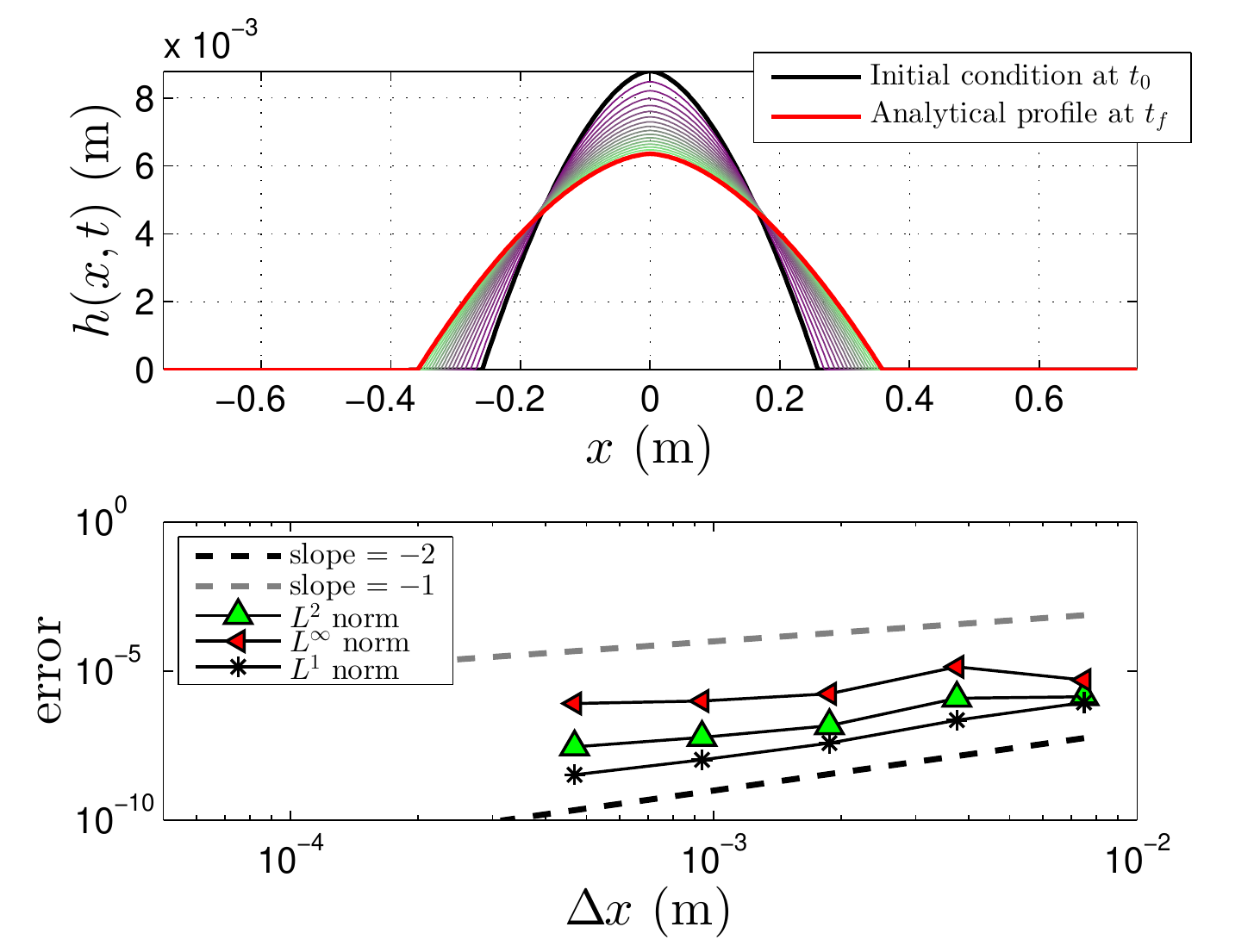}\label{fig:centsthin}}\hfill
\subfloat[][Shear-thickening fluid ($r=1.6$).]{\includegraphics[width=0.5\textwidth]{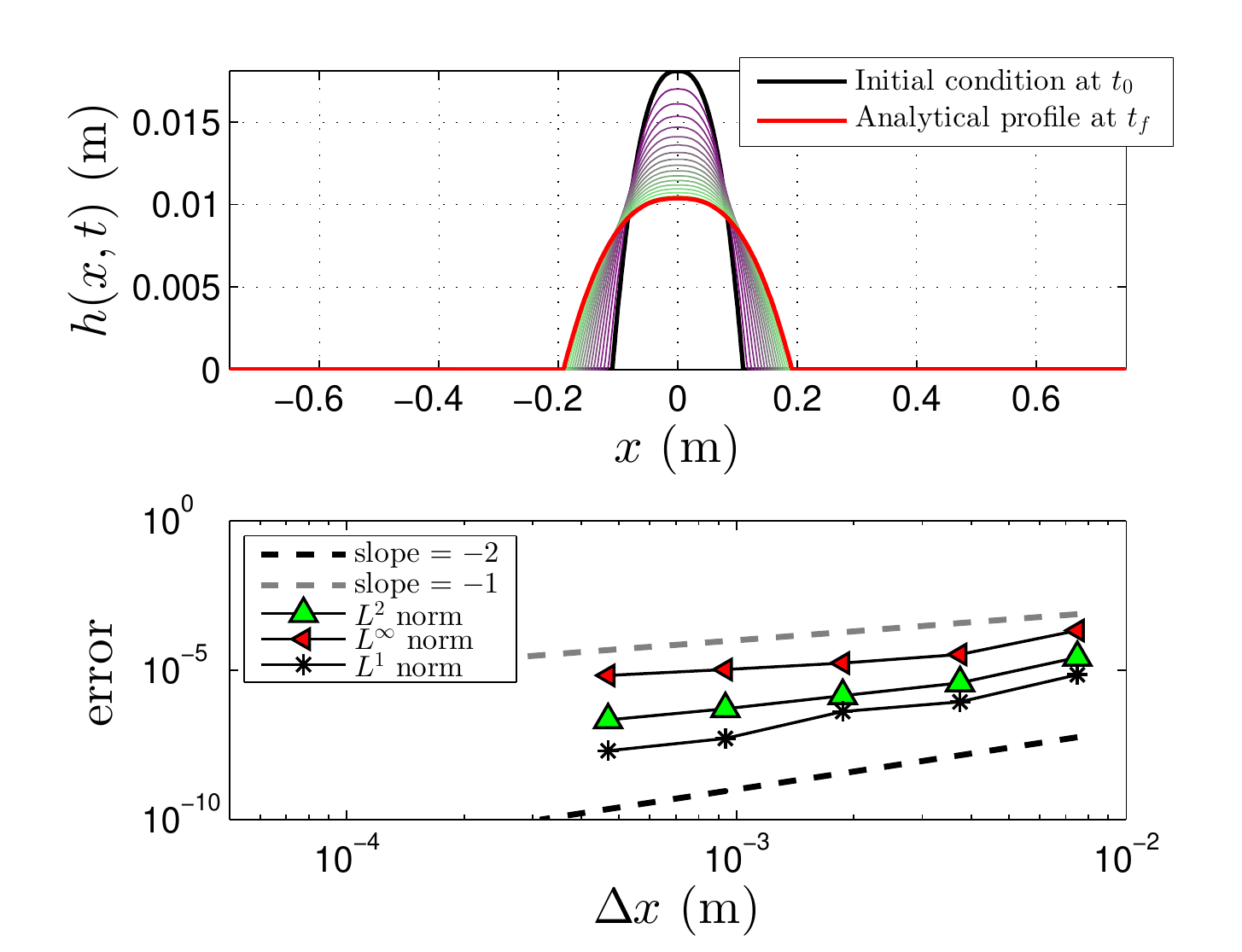}\label{fig:centsthick}}
\caption{Estimated order-of-convergence of a `centrally released' fixed fluid mass propagating in both directions in a uniform-width HS cell ($n=0$). The currents' shapes are plotted from `early' times (purple/dark) to `late' times (green/light). In all cases, the volume of fluid is $\mathcal{V}_0 = 2.4902 \times 10^{-5}$ m\textsuperscript{3} and $b_1 = 0.01739$ m. The currents are released at $t_0 = 1$ s and spread until $t_f = 3.5$ s.}
\label{fig:conv1}
\end{figure}

Figure \ref{fig:conv1} shows the propagation of constant-mass viscous gravity current of three different fluids: (a) Newtonian, (b) shear-thinning, and (c) shear-thickening power-law. The currents propagate symmetrically about the center of the domain ($x=0$). The sharp moving front $x_f(t)$ is accurately captured in these simulations on fairly modest (i.e., coarse) grids, without \emph{any} signs of numerical instability or need for special treatment of the derivative discontinuity. Computing the error as a function of $\Delta x$ during the grid refinement shows second-order convergence. This numerical example thus indicates that the proposed approach to treating the implicit nonlinear $\psi$ terms, specifically their evaluation at $n+1/2$, is consistent with the desired second-order accuracy. 

It should be noted that the restriction on $t_f$, which is necessary so that the current does not reach the domain boundaries, is critical since the chosen benchmark exact solution only describes the `spreading' behavior of the current and not its `levelling' (once it reaches the no-flux boundaries at $x=\pm L$). Indeed, the levelling regime possesses its distinct self-similar behavior (see, e.g., \cite{Diez1992,Zheng2018}), which is beyond the scope of the present work.

\subsubsection{Propagation of a Fixed Mass of Fluid in a Single Direction}
\label{sec:haf_conv}

To ascertain the accuracy of our discretization of the nonlinear BCs, we now return to a one-sided domain $x\in[\ell,L]$ with $\ell=0$. For the case of a current spreading away from the origin, the BC at the `left' end of the domain (from which the fluid is released) is non-trivial, and its proper discretization is key to the overall order of the convergence of the scheme. Conveniently, for a fixed mass ($\alpha=0$), the BCs still reduce to homogeneous Neumann conditions (recall \S\ref{sec:BC}), however, $h$ is no longer zero at the boundary (as was the case in \S\ref{sec:full_conv}). Thus, this benchmark is our successively `more complicated' case.

\begin{figure}[t]
\centering
\subfloat[][Newtonian fluid ($r=1$).]{\includegraphics[width=0.5\textwidth]{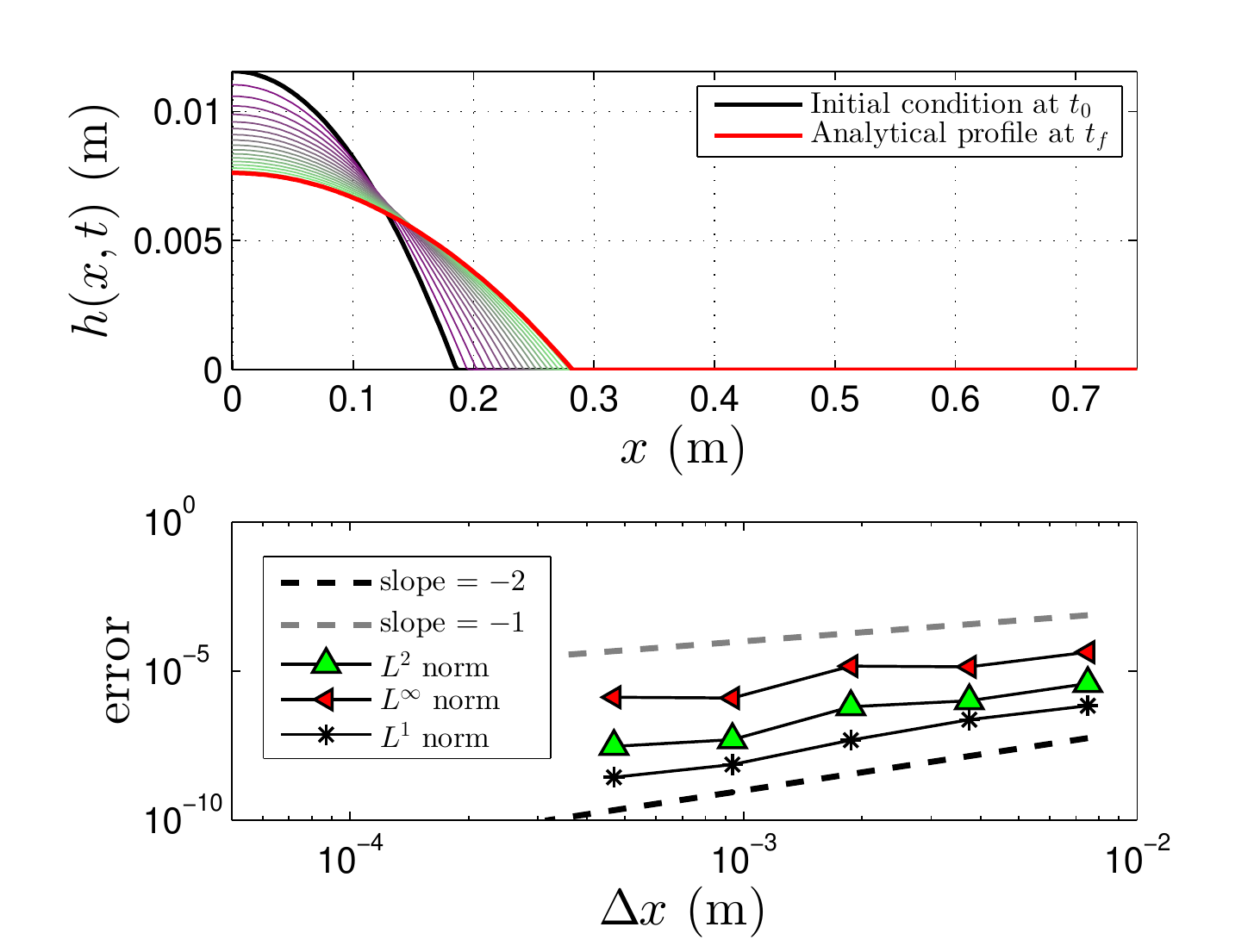}\label{fig:anaic_newt}}\\
\subfloat[][Shear-thinning fluid ($r=0.5$).]{\includegraphics[width=0.5\textwidth]{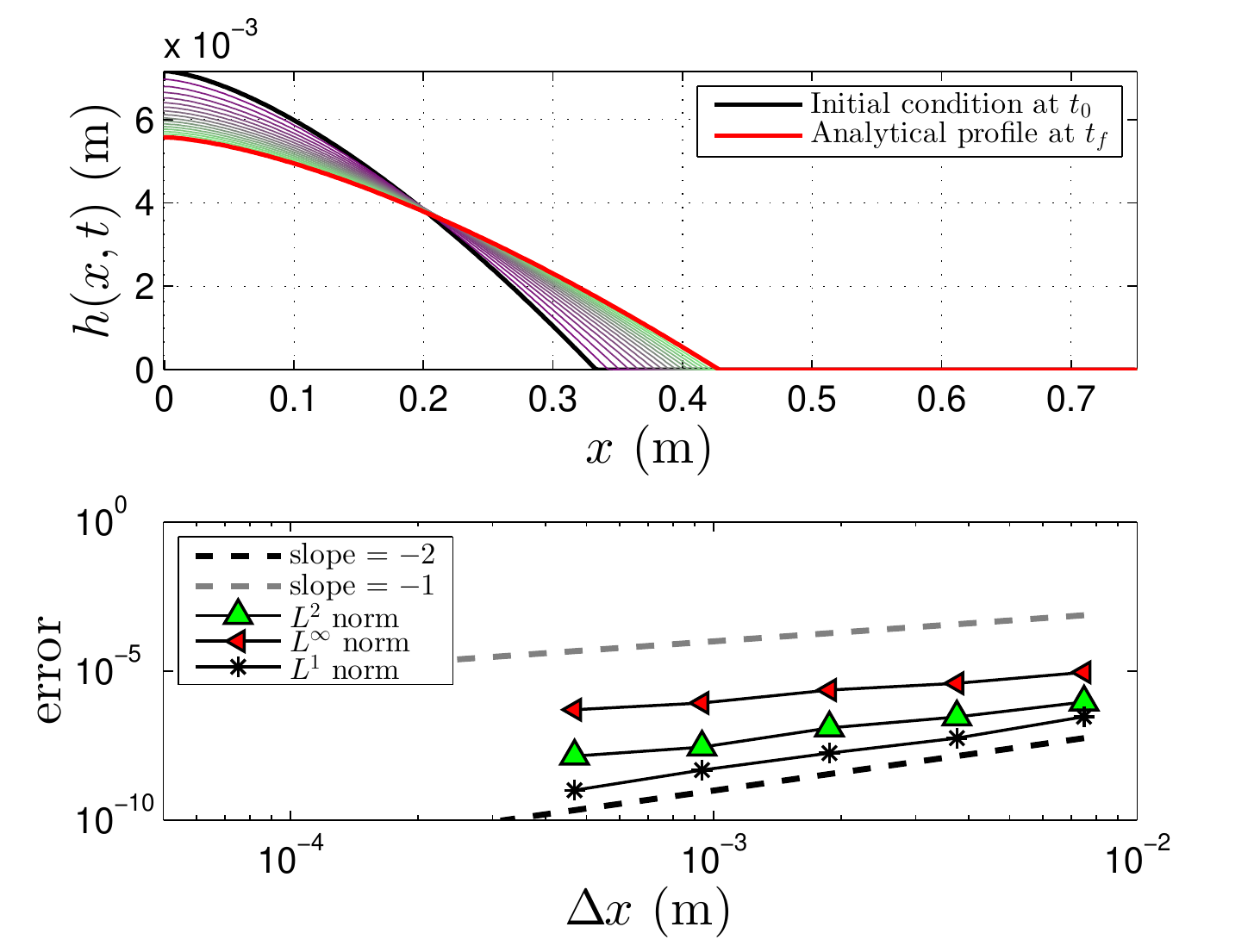}\label{fig:anaic_sthin}}\hfill
\subfloat[][Shear-thickening ($r=1.5$).]{\includegraphics[width=0.5\textwidth]{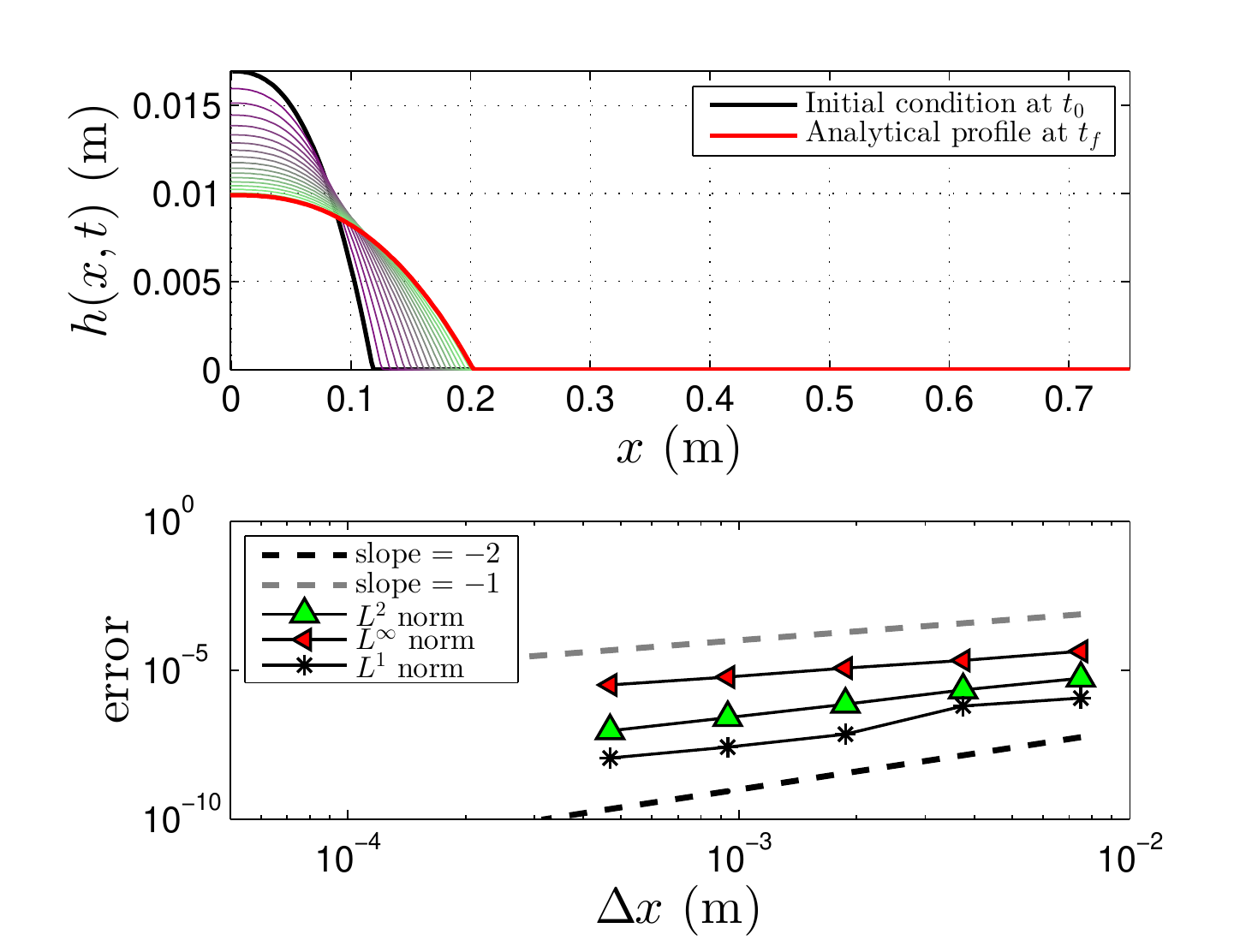}\label{fig:anaic_sthick}}
\caption{Estimated order-of-convergence study for the release of a fixed fluid mass propagating in a single direction (away from cell's origin) in a uniform-width HS cell ($n=0$). Once again, the fluid is released at $t_0 = 1$ s and spreads until $t_f = 3.5$ s. The currents' shapes are plotted from early times (purple/dark) through late times (green/light). The remaining model parameters for these simulations are the same as in figure~\ref{fig:conv1}.}
\label{fig:conv2}
\end{figure}

Once again, we ensure that $t_f$ is such that the fluid does not reach the downstream ($x=L$) domain end. Then, as in \S\ref{sec:full_conv}, we can once again use the exact solution of \citet{Ciriello2016} as the benchmark exact solution; again, this requires restricting to uniform-width HS cells (i.e., $n=0$).

Figure \ref{fig:conv2} shows clear second-order estimated order-of-convergence in the $L^1$ norm. This result indicates the decision to implement the scheme on a staggered grid, in which case the Neumann BCs (for $\alpha = 0$) are conveniently discretized using two-point central differences at the boundary, was indeed correct.

\subsubsection{Propagation in a Single Direction with Mass Injection}
\label{sec:haf_conv_inject}

Finally, we subject the numerical scheme to its most stringent test yet. That is, we compute the estimated order of convergence under mass injection conditions ($\alpha > 0$). The injection occurs near the cell's origin and the current propagates away from this location. Since $\alpha > 0$, the fully nonlinear forms of the  BCs as given in eqs.~\eqref{eq:bcs} and \eqref{eq:bcs2} now come into play.

Unlike the previously discussed cases of the release of a fixed fluid mass, a straightforward exact solution to the nonlinear ODE emerging from the self-similar analysis is not possible. For variable mass, obtaining a benchmark solution is now significantly more challenging, given that the nonlinear ODE must be solved \emph{numerically} (see the Appendix). Despite the availability of accurate stiff ODE solvers, such as {\tt ode15s} in {\sc Matlab}, it is quite difficult to map the numerical solution of the self-similar ODE onto the selected computational grid, and maintain the desired order of accuracy throughout this procedure. Therefore, for this benchmark, we instead elect to use a `fine-grid' numerical solution as the benchmark solution. This fine-grid solution is then compared against the solutions on successively coarser grids to establish the estimated order of convergence of the numerical scheme.

For this study, the simulation domain is $x \in [\ell,L]$ with $\ell = \Delta x_0$, so that $x_{i=0} = \ell - \Delta x_0/2$ and $x_{i=N} = L + \Delta x_0/2$; the boundary points are at the same cell faces on all grids during the refinement. The IC at $t_0 = 0$ s is from eq.~\eqref{eq:expontial_IC} with $b = 3.5 \times 10^2$, and $c = 25$ m\textsuperscript{$-1$}. 
The numerical solution is advanced up to $t_f = 1.5$ s.

\begin{figure}[t]
\subfloat[][$\alpha=1$]{\includegraphics[width=0.5\textwidth]{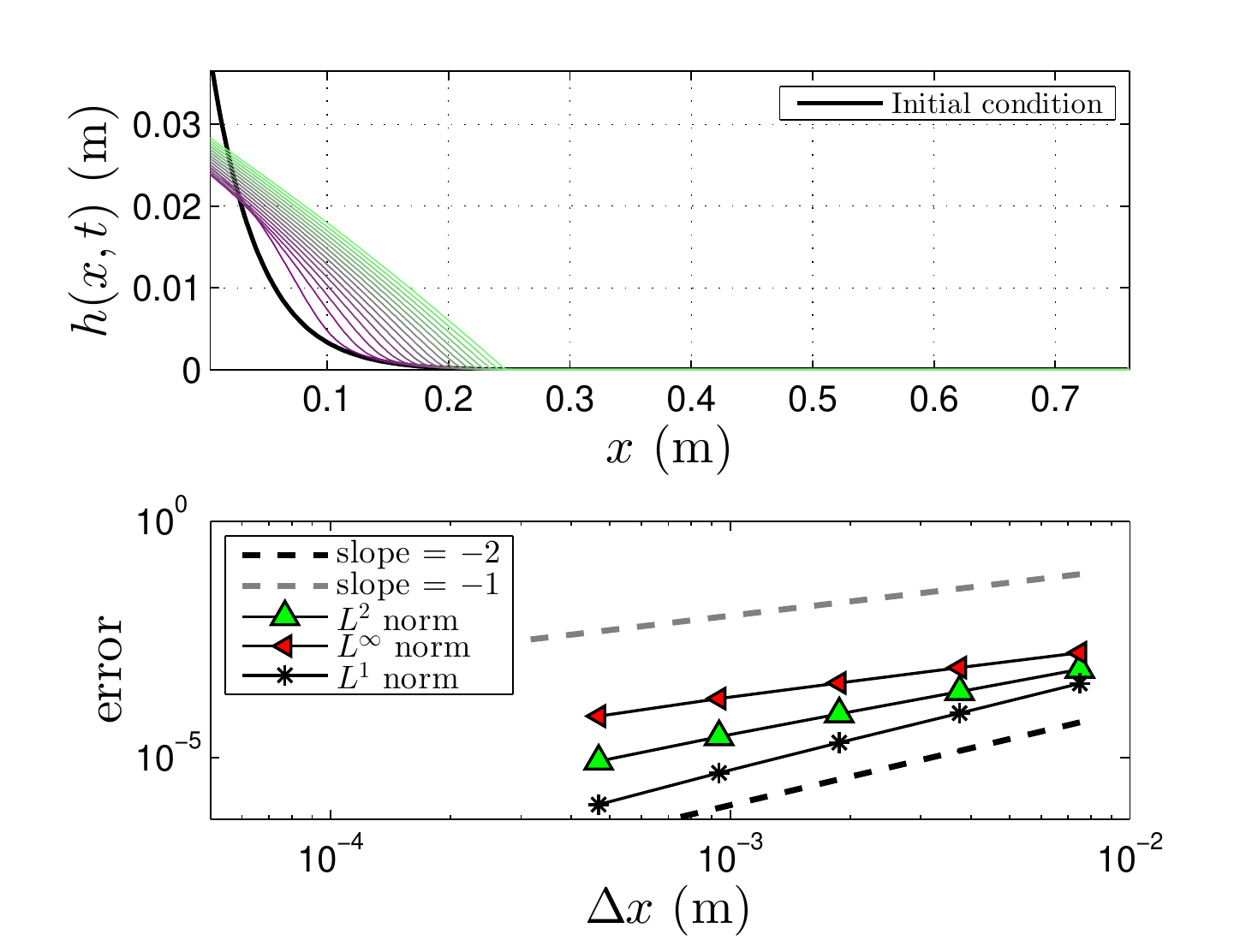}\label{fig:anaic_NBCnewt}}
\hfill
\subfloat[][$\alpha=1.5$]{\includegraphics[width=0.5\textwidth]{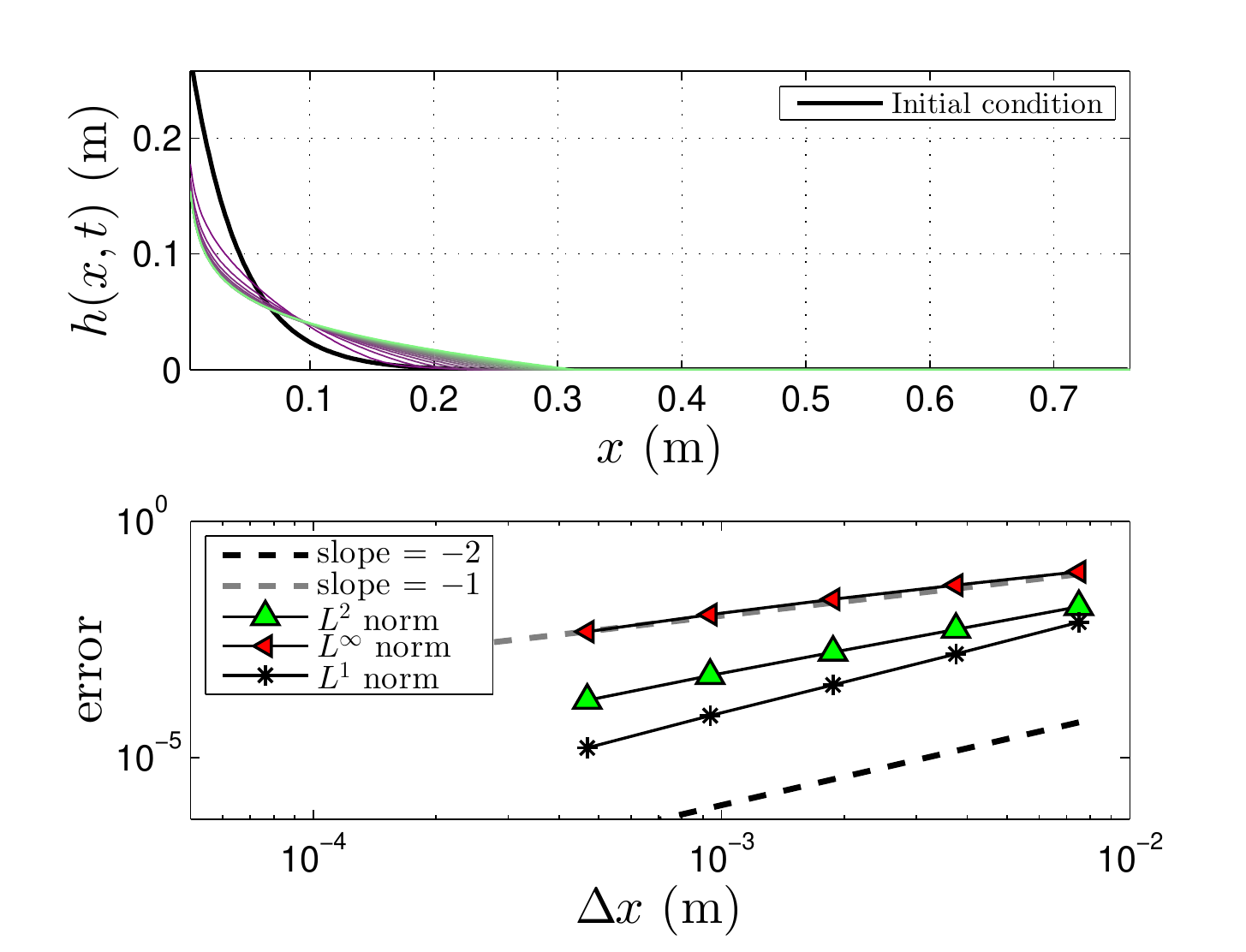}\label{fig:anaic_NBCnon_newt}}
\caption{Estimated order-of-convergence study for a variable-mass gravity current with injection at $x=\ell$ on the truncated domain $x\in[\ell,L]$ with $\ell=\Delta x_0$. Simulations are shown for the case of (a) a Newtonian fluid in a uniform-width HS cell ($r = 1$, $n = 0$), and (b) a shear-thickening fluid in a variable-width cell ($r=0.6$, $n = 0.6$). The remaining model parameters are as in figure~\ref{fig:conv1}.}
\label{fig:conv3}
\end{figure}

Figure \ref{fig:conv3} shows that the order of convergence of the numerical method is second order in space and time in the $L^1$ norm, as expected. In this figure, to show some variety, we present two distinct but arbitrarily selected cases: (a) a Newtonian fluid in a uniform-width HS cell ($n=0$) with volume growth exponent $\alpha = 1$, and (b) a non-Newtonian (shear-thinning, $r=0.6$) fluid in a variable-width HS cell (width exponent $n=0.6$) with a volume growth exponent $\alpha = 1.5$.

With this final numerical test, we have justified the formal truncation error of the proposed finite-difference scheme. This result is nontrivial because the PDE and the scheme are both \emph{nonlinear}, requiring subtle approximation on a staggered grid, across half-time steps, and linearization of an algebraic system via internal iterations.

\subsection{Satisfaction of the Mass Constraint at the Discrete Level}
\label{sec:consv}

Since the BCs derived in \S\ref{sec:eq} (and discretized in \S\ref{sec:BC}) stem from the mass conservation constraint (i.e., eq.~\eqref{eq:mass_conservn_porous} or eq.~\eqref{eq:mass_conservn}), it is expected that the proposed finite-difference scheme should produce a solution $h(x,t)$ that satisfies eq.~\eqref{eq:mass_conservn_porous} or eq.~\eqref{eq:mass_conservn} to within $\mathcal{O}\left[(\Delta x)^2+(\Delta t)^2\right]$ or better, if this constraint is checked independently after computing the numerical solution. To verify this capability of the scheme, in this subsection, we consider two cases: (i) a fixed fluid mass released near the origin ($\alpha = 0$), and (ii) spreading subject to mass injection ($\alpha > 0$) near the origin. Both cases are studied on the domain $x\in[\ell,L]$ with $\ell = \Delta x$. The solution is evolved on the time interval $t \in (t_0,t_f]$, and the volume error for each $t$ is computed as
\begin{equation}
     \left|\int_\ell^L h(x,t) b_1 x^n \,  \mathrm{d}x - (\mathcal{V}_0 +\mathcal{V}_\mathrm{in}t^\alpha)\right|,
    \label{mass_consv}
\end{equation}
where the $x$-integration is performed by the trapezoidal rule to $\mathcal{O}\left[(\Delta x)^2\right]$ on the staggered mesh, as before. We expect that the volume error, as defined in eq.~\eqref{mass_consv}, is $\mathcal{O}\left[(\Delta x)^2 + (\Delta t)^2\right]$, the same as the overall scheme. In this numerical study, the selection of the IC is no longer critical, as we do not compare against an exact self-similar solution. Accordingly, we select generic ICs from eqs.~\eqref{eq:poly_IC} and \eqref{eq:expontial_IC}.

\subsubsection{Fixed Mass Release ($\alpha=0$)}
\begin{figure}[hb]
\centering
\subfloat[][Newtonian fluid ($r=0$).]{\includegraphics[width=0.5\textwidth]{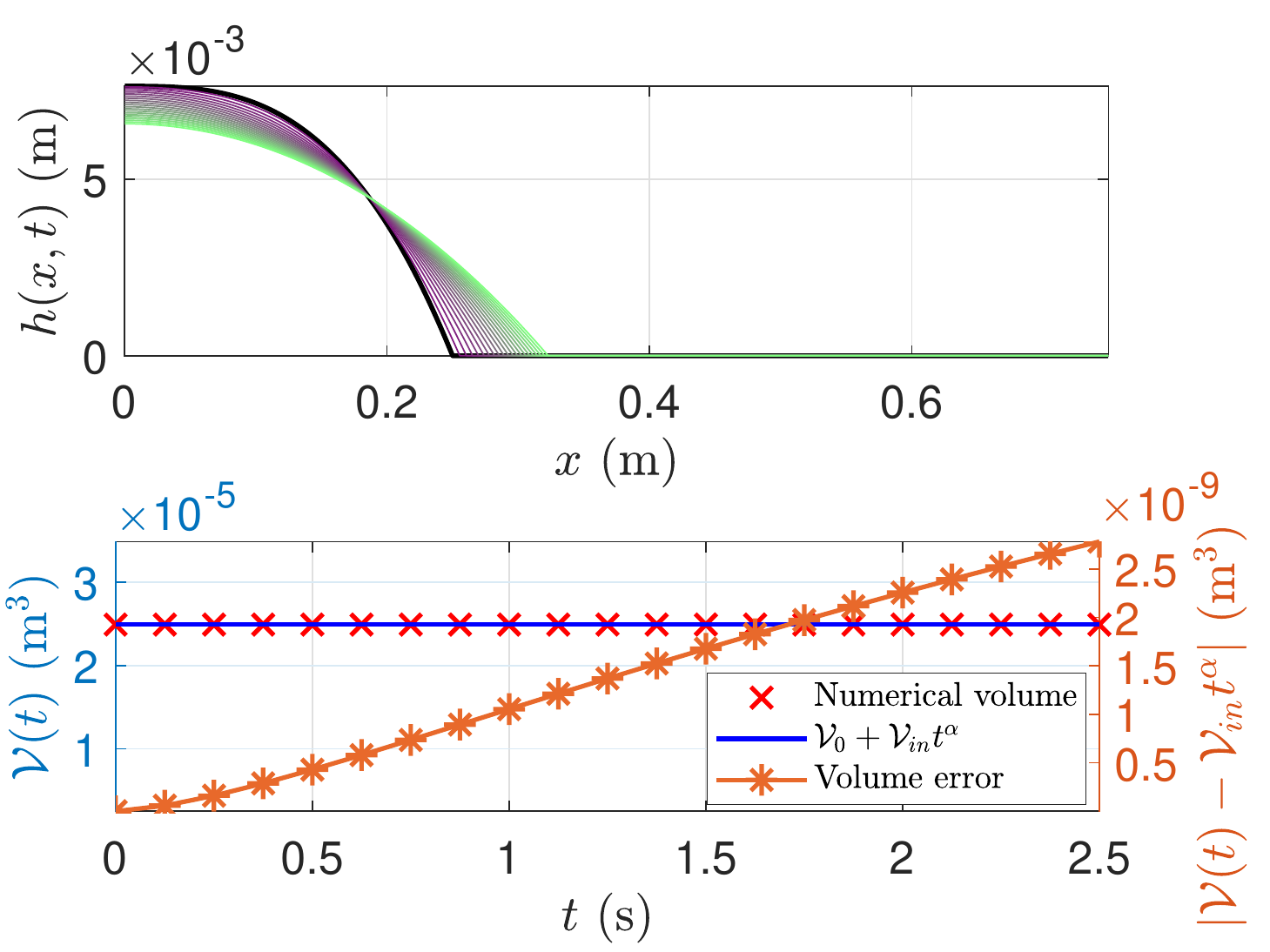}\label{fig:consvnewt}}\\
\subfloat[][Shear-thinning fluid ($r=0.7$).]{\includegraphics[width=0.5\textwidth]{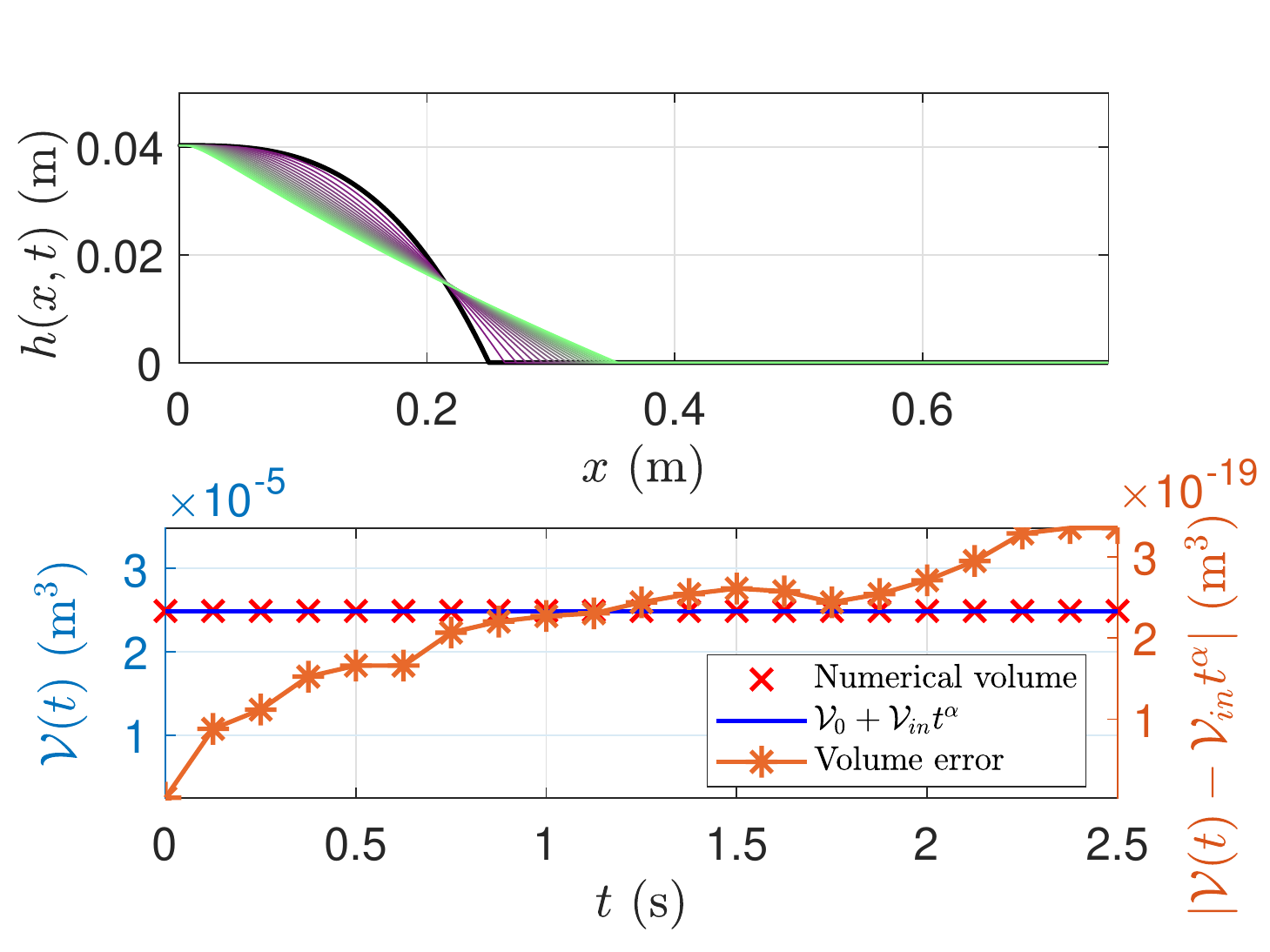}\label{fig:consvsthin}}\hfill
\subfloat[][Shear-thickening fluid ($r=1.5$).]{\includegraphics[width=0.5\textwidth]{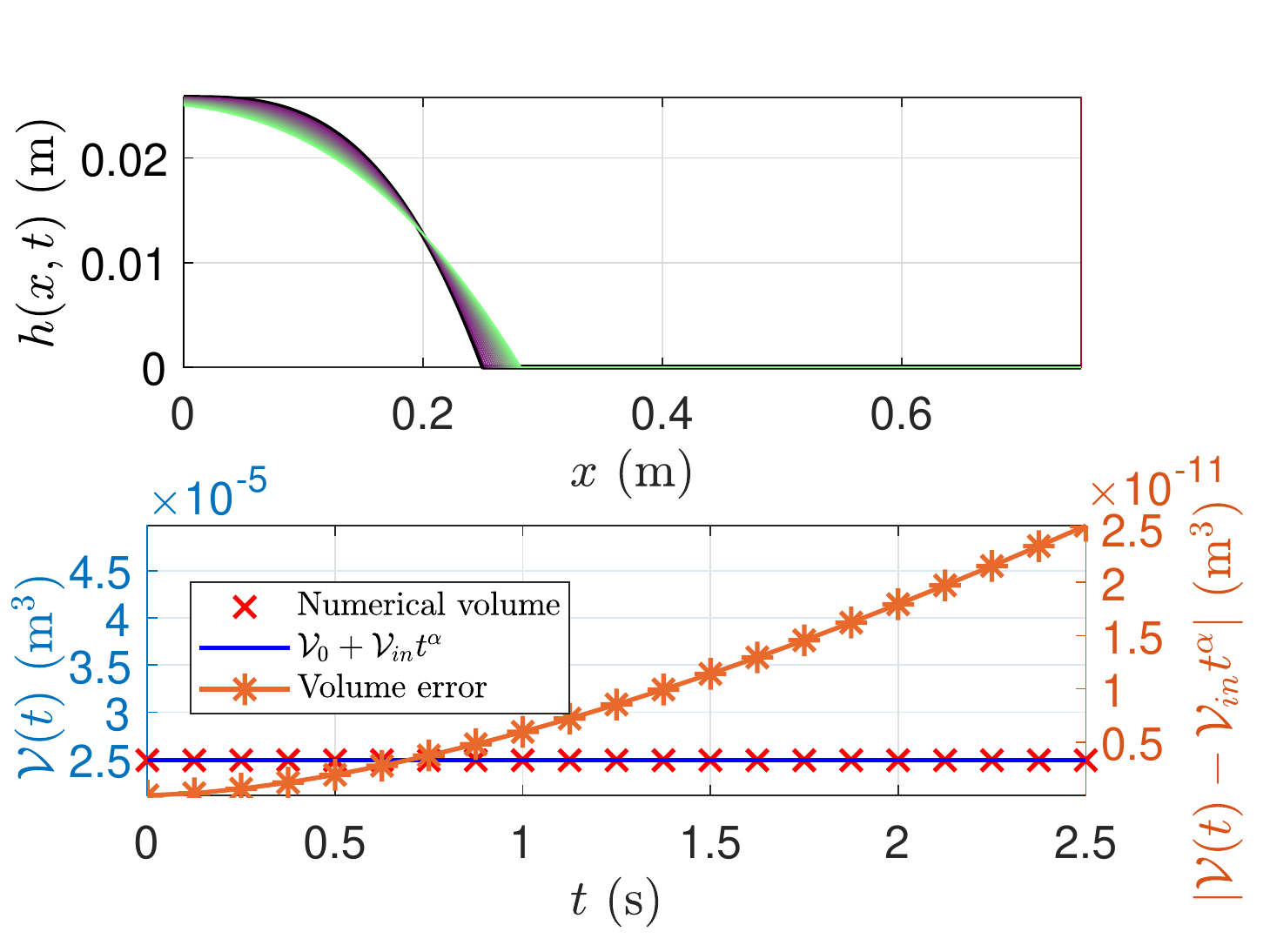}\label{fig:consvsthick}}
\caption{Results of the conservation study for the release of a fixed fluid mass. To highlight the scheme's capabilities, each case features a different HS cell: (a) $n=0$, (b) $n=0.7$, and (c) $n=0.5$. The currents are allowed to propagate from $t_0 = 0$ s up to $t_f = 2.5$ s, through 2500 time steps ($\Rightarrow \Delta t = 10^{-3}$ s). In all cases, $\alpha=0$ and  $\mathcal{V}(t) = \mathcal{V}_0 = 2.4902 \times 10^{-5}$ m\textsuperscript{3}. The remaining model parameters for these simulations are the same as in figure~\ref{fig:conv1}.}
\label{fig:cons_fixed_mass}
\end{figure}

In this case, the IC is a cubic polynomial determined from eq.~\eqref{eq:poly_IC} with $c=3$ and $\mathfrak{X}_0 = 0.25$ m. The error in the total fluid volume as a function of $t$ is compared with the initial one. Figure \ref{fig:cons_fixed_mass} shows that, while numerical error does build up in the total volume, the initial volume remains conserved to  within (or better than) $(\Delta t)^2=10^{-6}$.

\subsubsection{Mass Injection ($\alpha>0$)}
\begin{figure}
\centering
\subfloat[][$\alpha = 1$.]{\includegraphics[width=0.5\textwidth]{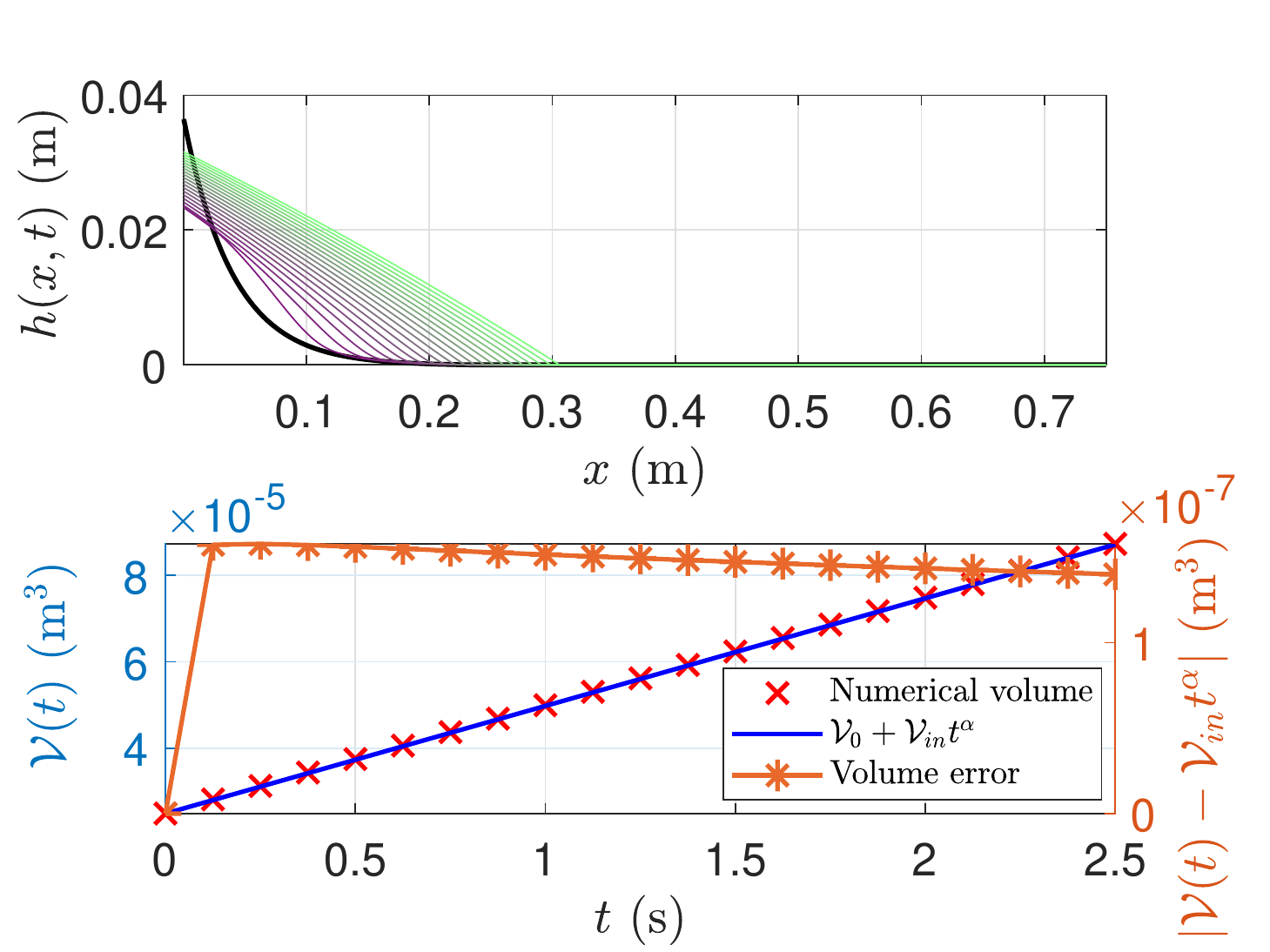}\label{fig:consvGnewt}}\\
\subfloat[][$\alpha = 1.5$.]{\includegraphics[width=0.5\textwidth]{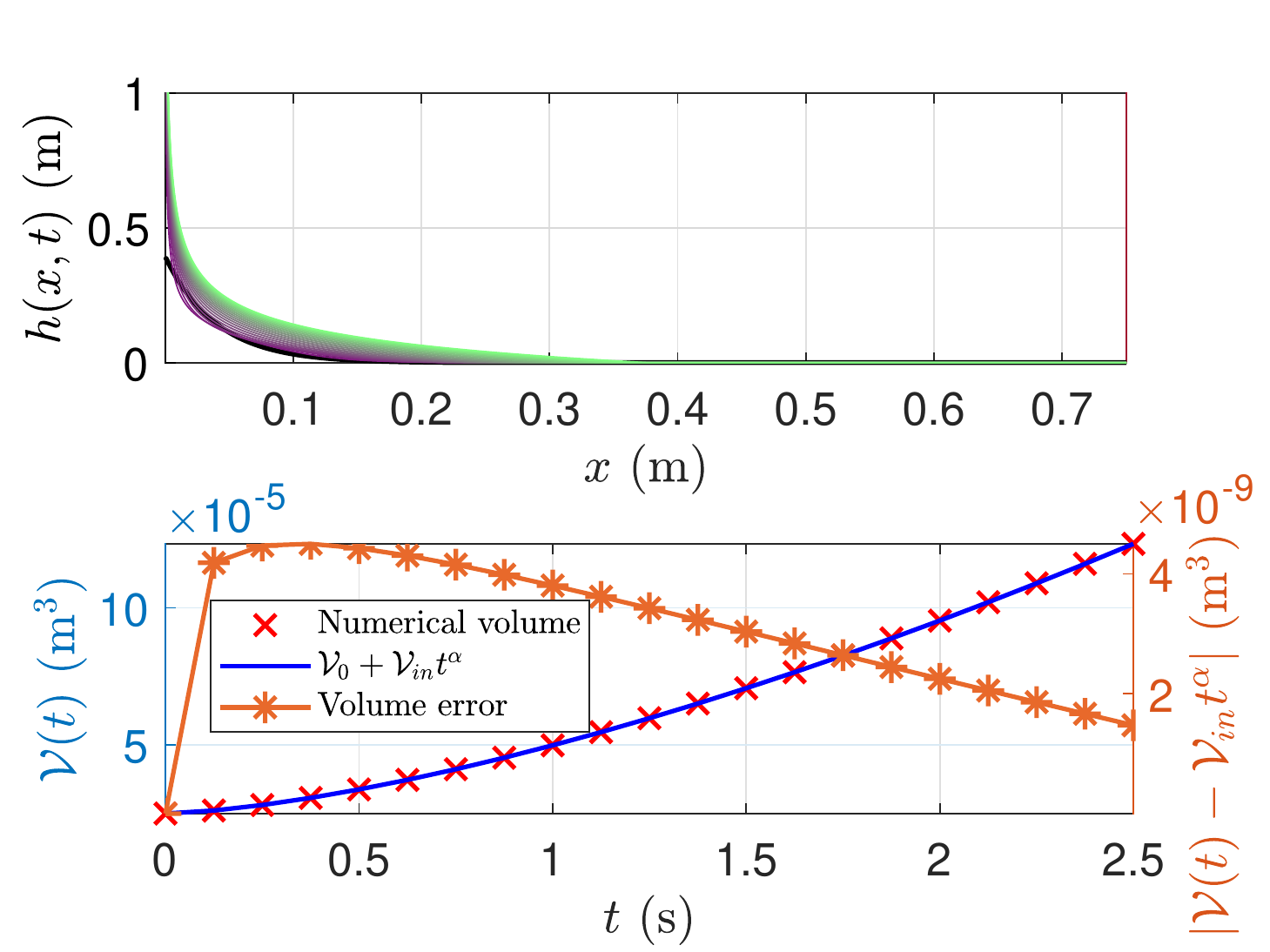}\label{fig:consvGsthin}}\hfill
\subfloat[][$\alpha=2$.]{\includegraphics[width=0.5\textwidth]{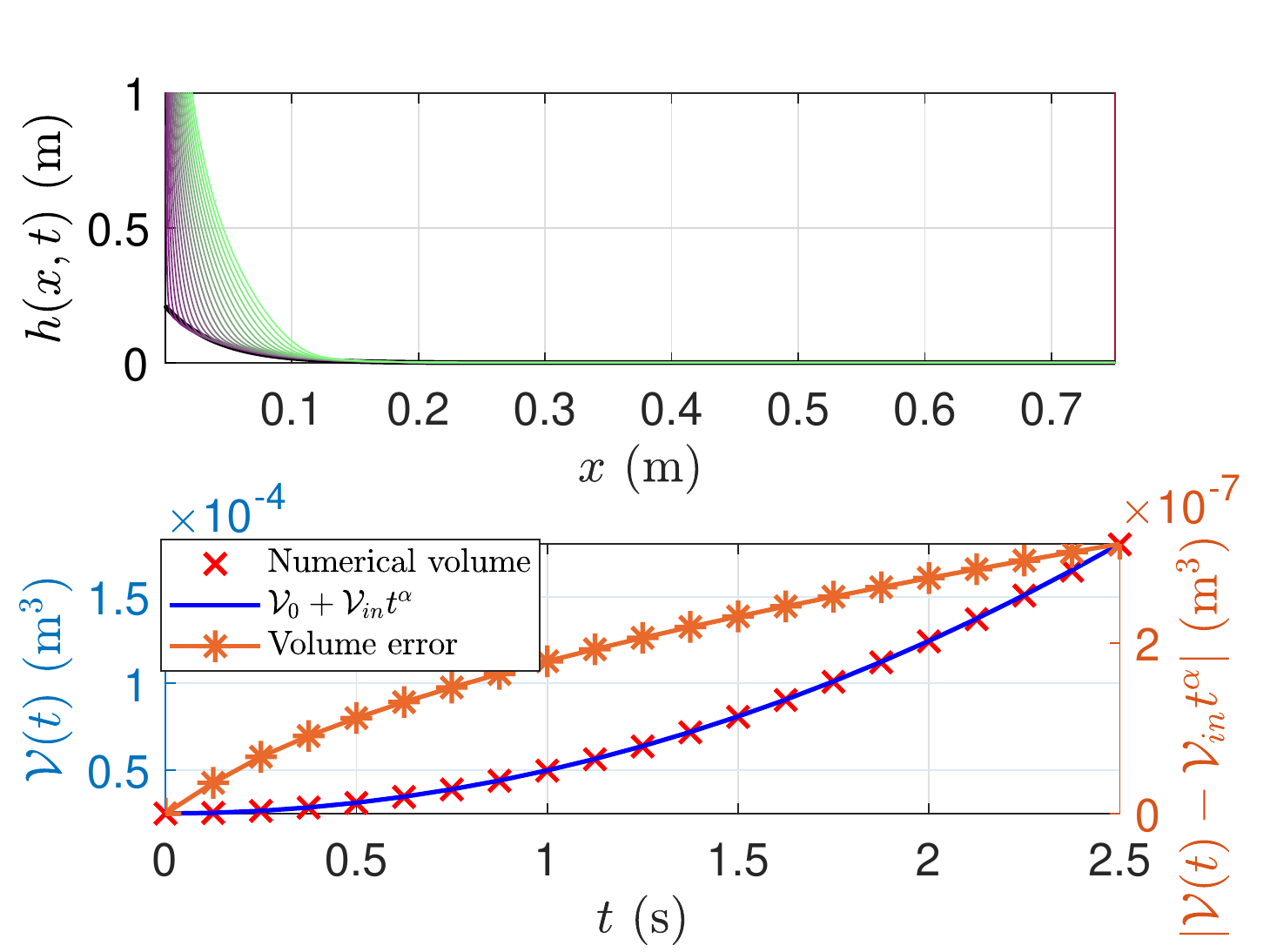}\label{fig:consvGsthick}}
\caption{Results of the conservation study for mass injection. Three choices of the volume exponent $\alpha$ are considered. We simulate (a) a Newtonian fluid in uniform HS cell ($r=1,n=0$), (b) a shear-thinning fluid in a variable-width HS cell ($r=0.7$, $n=0.7$), and (c) a shear-thickening fluid in a variable-width HS cell ($r = 1.5$, $n=0.5$). In all cases, $t_0$, $t_f$, and $\Delta t$ are as in figure~\ref{fig:cons_fixed_mass}. Additionally we set $\mathcal{V}_0 = \mathcal{V}_\mathrm{in} = 2.4902 \times 10^{-5}$ m\textsuperscript{3} (see eq.~\eqref{eq:mass_conservn}), The remaining model parameters for these simulations are the same as in figure~\ref{fig:conv1}.}
\label{fig:conservationMEG}
\end{figure}

A more stringent test of the conservation properties of the proposed scheme is conducted by applying the nonlinear BC associated with imposed mass injection at one end. For this case, the IC is taken to be the function in eq.~\eqref{eq:expontial_IC} with $b = 3.5\times10^{2}$ and $c = 25$ m$^{-1}$ and $\mathfrak{X}_0 = \frac{1}{c}\ln\frac{1}{b}$. A combination of  $n$, $r$ and $\alpha$ values have been considered to highlight the conservation properties across different physical regimes. Figure \ref{fig:conservationMEG} shows that, in all cases, the volume constraint is properly respected; while the volume error builds up, it remains small (within or better than $(\Delta t)^2=10^{-6}$).

\section{Conclusions and Outlook}

In this chapter, we developed and benchmarked a finite-difference numerical scheme for solving a family of nonlinear parabolic PDEs with variable coefficients given by eq.~\eqref{eq:nl_diff}. A special feature of these nonlinear PDEs is that they possess solutions that can propagate in a wave-like manner with a finite speed of propagation. Our study featured examples from this family of PDEs for modeling the 1D spreading (propagation) of a power-law (Oswald--de Weale) fluid in a horizontal, narrow fracture with variable width. We placed an emphasis on designing a series of numerical tests that show conclusively that the proposed scheme is second-order accurate in space and time. Analytical self-similar solutions for special cases of the nonlinear parabolic PDE were used to benchmark the numerical method. Furthermore, we verified that a global mass conservation/injection constraint can be successfully reformulated into a set of nonlinear boundary conditions, which were successfully discretized with second-order accuracy as well. 

The main advantage of the proposed finite-difference scheme is that it is strongly implicit, generalizing the time-stepping suggested by \citet{Crank1947}. Therefore, the proposed scheme does not formally require a time-step restriction for stability. By using a staggered grid, along the lines of \citet{Christov2009}, nonlinear terms were handled within the same three-point stencil as the classical Crank--Nicolson scheme. This choice of grid is particularly convenient for the discretization of the nonlinear boundary conditions, allowing second-order accuracy to be achieved with just a two-point stencil near the domain boundaries. Using fractional steps in time (`internal iterations'), we reformulated the nonlinear algebraic problem at each time step as a fixed-point iteration.

In future work, an interesting extension to our proposed numerical scheme could be the inclusion of a generic source term of the form $\mathcal{S}(x,t,h)$, added to the right-hand side of eq.~\eqref{eq:nl_diff}. Such a term can capture the effects of e.g., a leaky (porous) substrate over which a gravity current propagates, in which case $\mathcal{S}(x,t,h) = -\varkappa h(x,t)$ for some drainage constant $\varkappa$ \cite{Pritchard2001} (see also \cite[\S9.2]{Woods2015}). Then, the Crank--Nicolson disrectization in eq.~\eqref{eq:nl_diff_CNL} could be modified by adding 
\begin{equation}
    \frac{1}{2}\left[\mathcal{S}(x_i,t^{n+1},h_i^{n+1}) + \mathcal{S}(x_i,t^{n},h_i^{n}) \right]
\label{eq:source_term}
\end{equation}
to the right-hand side. Here, it is assumed that $\partial h/\partial x$ does not appear in $\mathcal{S}$ but only in $\mathcal{L}$. Therefore, the discretization in eq.~\eqref{eq:source_term} (even if nonlinear) will, at most, introduce a term in the matrix diagonal coefficient and a term on the right-hand side of eq.~\eqref{eq:fin_diff}. Another variation on this theme involves the spreading of an unconfined viscous fluid above a deep porous medium into which it penetrates in a time-dependent manner over a depth of $l(x,t)$ \cite{Acton2001}. Then, $\mathcal{S}(x,t,h) = -\kappa [ 1 + h(x,t)/l(x,t) ]$ and an additional ODE for $l(x,t)$ is coupled to eq.~\eqref{eq:nl_diff}. This problem is an interesting avenue for future extension of our proposed scheme, as we would have to discretize the ODE for $l(x,t)$ in the same Crank--Nicolson sense as eq.~\eqref{eq:nl_diff} and add an extra equation (row) to the discrete problem in eq.~\eqref{eq:fin_diff}.

On the other hand, an inclination angle (recall that all geometries in figure~\ref{fig:domains} are lying flat, so gravity is directed in the $-y$-direction) results in a term proportional to $\partial h/\partial x$ being added to eq.~\eqref{eq:nl_diff}  (for the case of a Newtonian fluid, see, e.g., \cite{Huppert1995,Vella2006}). This additional term changes the nonlinear diffusion equation~\eqref{eq:nl_diff} into a nonlinear \emph{advection--diffusion} equation. Care must be taken in discretizing this new advective term. A similar PDE arises in the segregation of bidisperse granular mixtures \cite{Dolgunin1995,Gray2015}. As discussed by \citet{ChristovMPM2018}, a strongly implicit Crank--Nicolson scheme can be successfully used for these problems. The scheme in \cite{ChristovMPM2018} is so robust that it performs well even in the singular vanishing-diffusivity limit of the advection-diffusion equation. Considering a generic advection term $\partial \Psi(h)/\partial x$, it can be handled analogously to the nonlinearity in the diffusion term. Specifically, we approximate
\begin{equation}
\left(\frac{\partial \Psi}{\partial x}\right)_{x=x_i} \approx \frac{1}{2}\left[ \left(\frac{\Psi_{i+1}^{n+1} - \Psi_{i-1}^{n+1}}{2 \Delta x}\right) + \left(\frac{\Psi_{i+1}^{n} - \Psi_{i-1}^{n}}{2 \Delta x}\right)\right].
\label{eq:adv_psi}
\end{equation}
Here, the advective term is discretized through a central difference formula involving a local three-point stencil on all interior nodes ($i=1$ to $N-1$). At the boundary nodes ($i=0$ and $i=N$), one can use a three-point biased (forward or backward) difference formula, as in eqs.~\eqref{eq:psi_boundary}. The now well-established idea of staggering the nonlinear term across fractional time steps is carried forward (recall eqs.~\eqref{eq:psi}). However, to properly linearize the advective term within the internal iterations, we must be able to write $\Psi(h) = \Upsilon(h)h$ (the most obvious way being to let $\Upsilon(h) \equiv \Psi(h)/h$) so that
\begin{equation}
\Psi_{i\pm1}^{n+1} \approx \Upsilon^{n+1/2,k}_{i\pm1} h^{n+1,k}_{i\pm 1}.
\label{eq:adv_psi_internal}
\end{equation}
Then, inserting eq.~\eqref{eq:adv_psi_internal} into eq.~\eqref{eq:adv_psi} and adding the result to the left-hand side of eq.~\eqref{eq:fin_diff}, modifies the tridiagonal system by adding $\Upsilon^{n+1/2,k}_{i\pm1}/(4\Delta x)$ to the superdiagonal ($i+1$) and subdiagonal ($i-1$), respectively. The remaining terms from eq.~\eqref{eq:adv_psi} are added to the right-hand side of the system.

Any of these potential extensions would have to be benchmarked against available first-kind self-similar solutions in \cite{Huppert1995, Acton2001, Vella2006, Woods2015}, however, no particular difficulties are expected to arise.

Another avenue of future work is as follows. Nowadays, high-order (i.e., greater than second-order) nonlinear parabolic PDEs are found to describe a wealth of low Reynolds number fluid phenomena: from the spreading and healing \cite{Zheng2018,ZhengHealing} to the rupture dynamics \cite{Garg2017} of thin liquid films dominated by capillary forces (see also \cite[Ch.~6-C]{L07}). Typically, the spatial operator is of fourth order due to the inclusion of surface tension effects (which depend upon the curvature of $h$), making the PDE more challenging to solve numerically. (Note that this is distinct from the inclusion of capillary effects in the context of gravity currents propagating in porous media, see \cite{Golding2011}.) Even higher (sixth) order thin film equations arise in dynamics of lubricated thin elastic membranes \cite{HM04,Flitton2004,Hewitt2015} dominated by elastic forces. To interrogate these complex interfacial phenomena, there is a current need for a robust and accurate numerical scheme to simulate these flows with low computational overhead (e.g., without the prohibitive time step stability restrictions of explicit schemes). In future work, it would be of interest to generalize the scheme from this chapter to such problems. Additionally, non-uniform (or adaptive) grids, which could be implemented along the lines of \cite{Christov2002}, can be used to capture singularity formation during thin film rupture.

\begin{acknowledgement}
We dedicate this work to the 80\textsuperscript{th} anniversary of Prof.\ J\"{u}ri Engelbrecht and his contributions to the fields of complexity science, nonlinear wave phenomena and applied mathematical modeling. I.C.C.\ acknowledges a very productive trip to Tallinn in September of 2014 (on invitation of Prof.~Andrus Salupere) and fondly recalls meeting and interacting with  Prof.~Engelbrecht there, at the IUTAM Symposium on Complexity of Nonlinear Waves.

We also thank Profs.~Tarmo Soomere and Arkadi Berezovski for their efforts in editing this volume and their kind invitation to contribute a chapter. Finally, I.C.C.\ acknowledges many helpful conversations on gravity currents with Prof.~H.A.\ Stone, Dr.~Z.\ Zheng and Prof.~S.\ Longo. Specifically, we thank S.\ Longo for suggesting the form of the governing equation for a power-law fluid in a variable-width HS cell (third row in table~\ref{tb:eq_models}).
\end{acknowledgement}

\section*{Appendix}

For many natural phenomena, a relatively simple procedure involving a scaling (dimensional analysis) of the governing equations can be used to yield the similarity variable and appropriate rescaling necessary to obtain a first-kind self-similar solution \cite{Barenblatt1979}. Viscous gravity currents exhibit such self-similar propagation, meaning that the solution (at sufficiently `long' times \cite{Barenblatt1979}) depends solely upon a combined variable of $x$ and $t$, rather than on each independently. Self-similarity allows for the derivation of exact analytical solutions to the governing equation~\eqref{eq:nl_diff} against which numerical solutions can be benchmarked. Specifically, for the case of the release of a fixed mass of fluid ($\alpha = 0$ so that $\mathcal{V}(t) = \mathcal{V}_\mathrm{in}$ $\forall t\in[t_0,t_f]$), a closed-form analytical self-similar solution was used in \S\ref{sec:conv} to test the order of convergence of the numerical scheme. 

In this Appendix, following \citet{DiFed2017}, we summarize the derivation of said self-similar solution for a power-law non-Newtonian fluid spreading \textit{away} from the origin ($x=0$) of a HS cell of uniform width $b_1$ ($n=0$).\footnote{While self-similar solutions, of course, exist for $n>0$, they cannot be found in closed-form as analytical solutions.} First, we introduce the following dimensionless variables (with $^*$ superscripts) from \cite{DiFed2017}: 
\begin{multline}
x^* = \left(\frac{B}{A^\alpha}\right)^{1/(\alpha-2)} x,\qquad t^* = \left(\frac{B}{A^2}\right)^{1/(\alpha-2)} t,\\
h^*(x^*,t^*) = \left(\frac{B}{A^\alpha}\right)^{1/(\alpha-2)} h(x,t),
\label{eq:scale}
\end{multline}
where $A$ is the constant from eq.~\eqref{eq:nl_diff} (defined in table \ref{tb:eq_models}) and $B = \mathcal{V}_\mathrm{in}/b_1$. Hereafter, we drop the  $^*$ superscripts. Next, we must select a suitable similarity variable $\eta$. As discussed in \S\ref{sec:intro}, a scaling analysis of the dimensionless version of eq.~\eqref{eq:nl_diff}, suggests that the self-similar solution of the first kind has the form
\begin{equation}
h(x,t) = \eta_N^{r+1}{t}^{F_2}f(\zeta), \qquad \zeta = \frac{\eta}{\eta_N}, \qquad \eta = \frac{x}{t^{F_1}}.
\label{eq:simsol}
\end{equation}
It can be shown that the constant $\eta_N$ specifically corresponds to the value of $\eta$ at the nose of the current, i.e., $\eta_N =  x_f(t)/{t}^{F_1}$, where $x=x_f(t)$ is such that $h\big(x_f(t),t\big) = 0$. Here, $\zeta$ is a convenient rescaled similarity variable, and the exponents $F_{1,2}$ are%
\begin{subequations}\begin{align}
    F_1 &= \frac{\alpha + r}{r+2}, \\
    F_2 &= \alpha - F_1.
\end{align}\label{eq:F1F2}\end{subequations}

The shape function $f(\zeta)$ represents the (universal) self-similar profile of the gravity current. We must now determine this function, by substituting eqs.~\eqref{eq:simsol} into the dimensionless version of eq.~\eqref{eq:nl_diff} to reduce the latter to a nonlinear ODE:
\begin{equation}
\frac{\mathrm{d}}{\mathrm{d} \zeta}\left(f\left|\frac{\mathrm{d} f}{\mathrm{d} \zeta}\right|^{\frac{1}{r}}\right) + F_2 f - F_1 \zeta \frac{\mathrm{d} f}{\mathrm{d} \zeta} = 0,\qquad \zeta\in[0,1].
\label{eq:nonlinODE}
\end{equation}
The second-order ODE in eq.~\eqref{eq:nonlinODE} can be rewritten as a first-order system:
\begin{equation}
\frac{\mathrm{d}}{\mathrm{d}\zeta}\left\lbrace
\begin{matrix}
f_1 \\
f_2
\end{matrix}
\right\rbrace
=
\left\lbrace
\begin{matrix}
f_2 \\
\displaystyle\frac{-r}{f_1 f_2 |f_2|^{(1-2r)/r}} \left[f_2|f_2|^{1/r} + F_2 f_1 - F_1 \zeta f_2\right]
\end{matrix}
\right\rbrace,
\label{eq:nonlinODE2}
\end{equation}
where, for convenience, we have set $f_1 = f$. The system in eq.~\eqref{eq:nonlinODE2} is `stiff,' and we must use an appropriate ODE solver, such as {\tt ode15s} in {\sc Matlab}, subject to appropriate initial and/or boundary conditions at $\zeta=0,1$.

A peculiarity of this self-similar analysis is that we have only a single BC for the ODE~\eqref{eq:nonlinODE}, namely $f(1)=0$, i.e., this is the location of the gravity current's nose $x=x_f(t)$ at which $\zeta=\eta/\eta_N=1$ and $h\big(x_f(t),t\big) = 0$. Since the ODE in eq.~\eqref{eq:nonlinODE2} requires a second initial or boundary condition, we use the `backwards-shooting' idea of \citet{Huppert1982a} to provide a second condition near $\zeta = 1$. Then, the ODE in eq.~\eqref{eq:nonlinODE2} can be integrated `backwards' from $\zeta = 1$ to $\zeta = 0$ subject to two `initial' conditions at $\zeta=1$. 

To this end, consider the asymptotic behavior of the current near the nose. By assuming that $f \sim \mathfrak{c}_1(1 - \zeta)^{\mathfrak{c}_2}$ as $\zeta\to1^-$ and substituting this expression into eq.~\eqref{eq:nonlinODE}, we obtain $\mathfrak{c}_1 = F_2^r$ and $\mathfrak{c}_2 = 1$ by balancing the lowest-order terms. Now, we have two BCs (see also \cite{DiFed2017}):%
\begin{subequations}\begin{align}
    f_1(1-\epsilon) &= F_2^r \epsilon,\\ 
    f_2(1-\epsilon) &= -F_2^r,
\end{align}\label{eq:f1f2_bc}\end{subequations}
for a sufficiently small $\epsilon \ll 1$. We can now solve the system~\eqref{eq:nonlinODE2} subject to the `final' conditions~\eqref{eq:f1f2_bc} on the interval $\zeta \in [0,1-\epsilon]$. By convention, an ODE is solved with initial, not final, conditions. Therefore, we perform the transformation $\zeta \mapsto 1-\hat{\zeta}$, which leads to the right-hand-side of eq.~\eqref{eq:nonlinODE2} being multiplied by $-1$. Then, the final conditions in eqs.~\eqref{eq:f1f2_bc} become initial conditions at $\hat{\zeta} = \epsilon$, and the first-oder system of ODEs is solved on the interval $\hat{\zeta} \in [\epsilon,1]$.

For certain special cases, a closed-form analytical solution to eq.~\eqref{eq:nonlinODE} can be obtained. For the case of the release of a fixed mass of fluid ($\alpha=0$), \citet{Ciriello2016} derived such an exact solution (as can be verified by substitution):
\begin{equation}
f(\zeta) = \frac{r^r}{(r+2)^r (r+1)} \left( 1 - \zeta ^{r+1} \right), 
\label{eq:unisol}
\end{equation}
which we used to benchmark our finite-difference scheme in \S\ref{sec:conv}. Finally, to obtain the viscous gravity current profile given in eq.~\eqref{eq:simsol} we must compute $\eta_N$. This value follows from imposing the mass conservation constraint in dimensionless form:
\begin{equation}
    \eta_N = \left[\int_0^1 f(\zeta) \,\mathrm{d} \zeta \right]^{-1/(r+2)} \approx \left[\int_\epsilon^1 f(\hat{\zeta}) \,\mathrm{d} \hat{\zeta} \right]^{-1/(r+2)},
    \label{eq:etaN}
\end{equation}
where the second (approximate) equality is needed for the case in which eq.~\eqref{eq:nonlinODE} has to be integrated numerically (no exact solution); $\epsilon \ll 1$ is chosen sufficiently small, as above. 

Finally we can substitute eqs.~\eqref{eq:unisol} and \eqref{eq:etaN} into eq.~\eqref{eq:simsol} to obtain the analytical solution for the profile of the gravity current, as a function of $x$ at some time $t$. It should be noted, however, that for this solution to apply, the current must have achieved its self-similar asymptotics , having forgotten the initial condition from which it evolved.


\footnotesize{
\bibliographystyle{spbasic}
\bibliography{AWM2_references.bib}
}

\end{document}